\documentclass{IEEEtran}

\usepackage{mypreamble}

\usepackage{lipsum}

\newcommand{\maketextstyle}{\textstyle} %

\begin{document}

\title{Nonlinear Bode Diagrams for MIMO Systems via IQCs and the Generalized KYP Lemma}

\author{Julius P.~J. Krebbekx$^1$, Roland Tóth$^{1,2}$, Amritam Das$^1$
\thanks{$^1$Control Systems group, Department of Electrical Engineering, Eindhoven University of Technology, 5600 MB Eindhoven, The Netherlands.}
\thanks{$^2$Systems and Control Lab, HUN-REN Institute for Computer Science and Control, 1111 Budapest, Hungary.}
\thanks{E-mail: {\tt\small \{j.p.j.krebbekx, r.toth, am.das\}@tue.nl}}
}

\date{\today}

\maketitle

\begin{abstract}
    The Bode diagram forms the foundation of performance analysis and shaping for MIMO LTI systems. In this paper, we develop a graphical frequency-domain analysis tool that generalizes the Bode diagram to nonlinear systems that preserve the period of the input in the output. We consider three spaces of periodic signals as function of frequency: pure sinusoids, signals with higher harmonics, and signals with bounded frequency content. The nonlinear Bode plot is obtained by plotting the $L_2$-gain restricted to these spaces as function of frequency. We use the IQC framework to derive \emph{Frequency-Domain Inequalities} (FDIs) that certify $L_2$-gain performance on each of the three input spaces, and subsequently formulate those FDIs as semidefinite programs via the generalized KYP lemma. Finally, we demonstrate the effectiveness of our proposed methods on a controlled unbalanced disk setup. For this example, we show how the Bode diagram is computed in practice, and that Zames-Falb multipliers can be used to correctly quantify the closed-loop integrator behavior in our Bode diagram. Additionally, the performance described by the nonlinear Bode diagram is verified in model-based simulations and experiments on the setup.
\end{abstract}

\begin{IEEEkeywords}
Frequency-domain analysis, Nonlinear control systems, Integral quadratic constraints, MIMO systems %
\end{IEEEkeywords}

\section{Introduction}

\IEEEPARstart{F}{or} \emph{Linear Time-Invariant} (LTI) systems, the Nyquist\cite{nyquistRegenerationTheory1932} and Bode~\cite{bodeRelationsAttenuationPhase1940} diagrams form the foundation of modern control engineering. These graphical frequency-domain methods allow one to specify, analyze and shape closed-loop performance in a frequency-wise manner, e.g., via weighting filters in mixed-sensitivity design~\cite{skogestadMultivariableFeedbackControl2010,zhouRobustOptimalControl1996}. When nonlinear dynamics start to play a role, tools based on the LTI assumption no longer provide reliable performance guarantees. 

Extending these graphical frequency-domain performance tools to the nonlinear setting has been the subject of extensive research for several decades, dating back to the \emph{Describing Function} (DF)~\cite{krylovIntroductionNonlinearMechanics1943}. While the DF has been applied successfully in practice~\cite{pogromskyFrequencyDomainPerformance2014,heertjesVariableGainMotion2016}, it is approximative by nature and provides no guarantees. Non-approximative methods typically rely on the property that a periodic input yields a unique periodic steady-state output, which is guaranteed for convergent~\cite{pavlovUniformOutputRegulation2006}, contracting~\cite{lohmillerContractionAnalysisNonlinear1998}, or fading memory~\cite{boydFadingMemoryProblem1985} systems, to name a few. For convergent systems, frequency-domain performance has been quantified for sinusoidal inputs~\cite{pavlovFrequencyDomainPerformance2007}. Other works consider higher harmonics in the input, see~\cite{rijlaarsdamComparativeOverviewFrequency2017} and references therein. Alternatively, one can consider a specific periodic input signal, in which case the steady-state solution is computed explicitly~\cite{moreschiniFrequencyResponseNonlinear2026,pavlovFastComputationFrequency2008,pavlovSteadystatePerformanceOptimization2013}. All approaches mentioned above are either approximative, restricted to specific inputs, or require direct computation of the steady-state response itself.

In our previous work~\cite{krebbekxNonlinearBandwidthBode2025a}, we defined a notion of frequency-dependent $L_2$-gain by restricting the input to certain classes of periodic signals. This led to a nonlinear Bode diagram for \emph{Single-Input Single-Output} (SISO) systems, for which the computation is based on \emph{Scaled Relative Graphs} (SRGs)~\cite{ryuScaledRelativeGraphs2022,chaffeyGraphicalNonlinearSystem2023}. While SRGs are graphical and intuitive, the SRG-based computation of the Bode diagram requires gridding and numerical complex geometry calculations for every frequency. The extension to \emph{Multiple-Input Multiple-Output} (MIMO) systems~\cite{krebbekxMIMOPaperExtended2026} with multiple nonlinearities is not straightforward for this purpose, and can lead to conservative bounds.

In this paper, we define a notion of frequency-dependent $L_2$-gain for nonlinear and/or uncertain MIMO systems, and provide a computationally tractable toolchain via the framework of \emph{Integral Quadratic Constraints} (IQCs)~\cite{megretskiSystemAnalysisIntegral1997}. As in~\cite{krebbekxNonlinearBandwidthBode2025a}, we define three spaces of periodic inputs at a given frequency: the sinusoidal, harmonic, and subharmonic inputs. The sinusoidal inputs are all sinusoids with a given frequency, and the harmonic space contains all higher harmonics as well. The subharmonic space contains all periodic signals with frequency content up to a given frequency, which is a novel concept and can be used to analyze low-frequency behavior. For nonlinear systems that are \emph{period-preserving}\footnote{Systems that have a unique steady-state response with the same period as the input. See Definition~\ref{def:period-preserving} for a formal definition.}, the sinusoidal, harmonic and subharmonic gains are obtained by computing the $L_2$-gain of a nonlinear system with inputs restricted to the respective input space. For period-preserving systems, we prove that the time-domain $L_2$-gain from periodic input to output, including transient effects, asymptotically equals the \emph{Root Mean Square} (RMS) gain from periodic input to steady-state output, which strengthens~\cite[Lemma 1]{krebbekxNonlinearBandwidthBode2025a}. The nonlinear Bode diagram is obtained by plotting the frequency-dependent gains as function of frequency for each SISO transfer, with an additional column that captures the combined effect of all inputs on one output via an input filter. Our approach goes beyond the existing methods since it covers all periodic inputs, not just sinusoids or specific inputs, and it is not approximative since it yields guaranteed upper bounds for the frequency-dependent $L_2$-gain. Moreover, this paper generalizes~\cite{krebbekxNonlinearBandwidthBode2025a} to the MIMO setting with multiple nonlinearities.

We then turn to the computation of the frequency-dependent sinusoidal, harmonic and subharmonic gains using IQCs. In the IQC framework, nonlinear and/or uncertain components are described by multipliers, and stability and performance of an interconnection are verified via a \emph{Frequency-Domain Inequality} (FDI)~\cite{jonssonLectureNotesIntegral2001,veenmanRobustStabilityPerformance2016,schererDissipativityIntegralQuadratic2022}. If the multipliers admit a finite-dimensional realization, the FDI can be verified over all frequencies using the \emph{Kalman-Yakubovich-Popov} (KYP) lemma~\cite{rantzerKalmanYakubovichPopov1996}, and over specific frequency ranges via the \emph{Generalized KYP} (GKYP) lemma~\cite{iwasakiGeneralizedKYPLemma2005}, both yielding a convex \emph{Linear Matrix Inequality} (LMI) feasibility problem. As IQC analysis treats MIMO systems naturally, allows for a refined description of nonlinearities through various multipliers~\cite{veenmanRobustStabilityPerformance2016}, and is solved via semidefinite programming, it offers significant potential advantages over the SRG approach from~\cite{krebbekxNonlinearBandwidthBode2025a} in terms of conservatism and computational speed. 

We use the IQC framework to derive FDIs involving specific frequency ranges that certify performance w.r.t. a quadratic supply rate for inputs in the sinusoidal, harmonic, and subharmonic spaces. We subsequently formulate semidefinite programs that use the GKYP lemma to verify our FDIs for quadratic performance on the signal spaces of interest, yielding upper bounds for the frequency-dependent gains that make up the nonlinear Bode diagram. A related approach was used in~\cite{jonssonAnalysisPeriodicallyForced2003} for square-integrable periodic inputs. In~\cite{jonssonAnalysisPeriodicallyForced2003}, however, no further refinement of the frequency content of the input was made, and different performance objectives were considered. Also, their approach results either in a non-smooth optimization problem, or resorts to truncation of the FDI to a finite number of harmonics.

Finally, we apply our theoretical results for the performance analysis of a controlled unbalanced disk. Via this example, we demonstrate how the nonlinear Bode diagrams are computed in practice, and how they can be used to quantify the performance of different controllers. For a PID controller, we show that \emph{Zames-Falb} (ZF) multipliers~\cite{zamesStabilityConditionsSystems1968} accurately quantify the integrator behavior of the closed loop. We validate the performance predicted by the Bode diagram via simulations, and perform the same experiments on the setup to show that our approach is not critically sensitive to modelling errors. 

In summary, the contributions of this paper are:
\begin{itemize}
    \item \emph{The Nonlinear Bode Diagram} (Section~\ref{sec:gammas_bode_def}): We define the sinusoidal, harmonic and subharmonic signal spaces and corresponding gains. Then, we develop an improved version of~\cite[Lemma 1]{krebbekxNonlinearBandwidthBode2025a} to bridge the time-domain and steady-state performance of a period-preserving system, and define the nonlinear Bode diagram for MIMO systems with multiple nonlinearities, generalizing~\cite{krebbekxNonlinearBandwidthBode2025a}.
    \item \emph{Computation via IQCs} (Section~\ref{sec:nl_bode_computation}): For period-preserving systems with inputs in the sinusoidal, harmonic, or subharmonic space, we derive FDIs that guarantee $L_2$-gain performance, and formulate semidefinite programs to verify these FDIs.
    \item \emph{Application} (Section~\ref{sec:application}): %
    We apply our proposed methods on the performance analysis of a controlled unbalanced disk setup, which includes computation of the Bode diagrams, and validation via simulations and experiment.
\end{itemize}

Moreover, we present the preliminaries in Section~\ref{sec:preliminaries}, and our conclusions in Section~\ref{sec:conclusion}.

\section{Preliminaries}\label{sec:preliminaries}

\subsection{Notation and Conventions}

Let $\N$ ($\N_+$) denote the (nonzero) natural numbers, $\Z$ the integers, and $\R,\C$ the real and complex number fields, respectively, and $\R_{\geq 0} := [0,\infty)$. We denote the complex conjugate of $z = a + jb \in \C^n$ as $\bar{z} = a-jb$, where $j^2=-1$, $a,b \in \R^n$ and $n \in \N_+$. The norm on $\C^n$ (and $\R^n$) is defined as $|z| := \sqrt{z^* z}$, where $z^* := \bar{z}^\top$ is the conjugate transpose. %
Let $\mathbb{S}^n$ denote the real symmetric $n \times n$ matrices, and $I$ denotes the appropriately sized identity matrix. Expressions $G^* P G$, where $P^*=P$, are often abbreviated as $(\star)^* P G$.

\subsection{Signals}

For $d \in \N_+$ and $T >0 $, we consider the Hilbert spaces $L_2^d := \{u : \R_{\geq 0} \to \R^d \mid \norm{u} < \infty \}$ with $\norm{u}^2 := \inner{u}{u}$ and $\inner{u}{v} := \int_0^\infty u(\tau)^\top v(\tau) \dd \tau$, and $L_2^d[0,T] := \{u : [0,T] \to \R^d \mid \norm{u}_{T} < \infty \}$ with $\norm{u}_T^2 := \inner{u}{u}_T$ and $\inner{u}{v}_T := \int_0^T u(\tau)^\top v(\tau) \dd \tau$. Elements of $u \in L_2^d$ (or $L_2^d[0,T]$) are indexed as $u=(u_1 \dots u_d)^\top$, with $u_i \in L_2$ (or $L_2[0,T]$).%

For $T \in \R_{\geq 0} $, define the truncation operator $P_T$ on signals $u: \R_{\geq 0} \to \R^d$ as $(P_T u)(t) = 0$ for all $t>T$, else $(P_T u)(t) = u(t)$. The extended $L_2$ space is defined as $\Lte^d := \{u : \R_{\geq 0} \to \R^d \mid \norm{P_T u} < \infty, \, \forall T \in \R_{\geq 0} \}$~\cite[p. 38]{desoerFeedbackSystemsInputoutput1975}, and is useful as it includes periodic, persistent and unbounded signals, which are excluded from $L_2$.

A signal $u : \R_{\geq 0} \to \R^d$ is called $T$-periodic for some $T>0$, if $u(t+T)=u(t)$ for all $t \geq 0$. Any $T$-periodic signal $u\in \Lte^d$ can be uniquely viewed as an element of $v \in L_2^d[0,T]$ via $v = P_T u$, and vice versa. Any $u \in L_2^d[0,T]$ can be written in Fourier series as $u(t) = \sum_{n \in \Z} \hat{u}_n e^{2 \pi j n t/T}$ where $\hat{u}_n \in \C^d$ are the Fourier coefficients $\hat{u}_n = \frac{1}{T} \int_0^T u(\tau) e^{-2 \pi j n \tau/T} \dd \tau$. The RMS norm of the signal is defined as
\begin{equation}\label{eq:rms_fourier_L2}
    \maketextstyle \norm{u}_\mathrm{RMS} := \sqrt{\sum_{n \in \Z} |\hat{u}_n|^2} = 1/\sqrt{T} \norm{u}_{T}.
\end{equation}

\subsection{Systems}

A system with $p$ inputs and $q$ outputs is modeled as an operator $R : \Lte^p \to \Lte^q$. A system is \emph{causal} if it satisfies $P_T (R w) = P_T(R(P_T w))$, i.e., the output at time $T$ is independent of the input at times greater than $T$. 

We assume that systems obey $R0=0$. Then, the $L_2$-gain of a system $R : \Lte^p \to \Lte^q$ is defined as
\begin{equation}\label{eq:L2-gain}
    \maketextstyle \gamma(R) := \sup_{0 \neq w \in L_2^p} \frac{\norm{Rw}}{\norm{w}} \in [0,\infty].
\end{equation}
A system $R$ is called stable if $\gamma(R) < \infty$. Similarly, the incremental $L_2$-gain is defined as
\begin{equation}\label{eq:incr-L2-gain}
    \maketextstyle \Gamma(R) := \sup_{w_1,w_2 \in L_2^p, w_1 \neq w_2} \frac{\norm{Rw_1-Rw_2}}{\norm{w_1-w_2}} \in [0,\infty].
\end{equation}

\begin{figure}[t]
    \centering

    \begin{subfigure}[b]{\linewidth}
        \centering
        \begin{tikzpicture}
            \draw[black] (-.4,-.4) rectangle (.4,.4);
            \node at (0,0) {$P$};

            \draw[black] (-.3,-.3+.9) rectangle (.3,.3+.9);
            \node at (0,.9) {$\Phi$};

            \draw[<-] (-1,-.2) -- (-.4,-.2);
            \node at (-1.2,-.2) {$z$};

            \draw[<-] (.4,-.2) -- (1,-.2);
            \node at (1.2,-.2) {$w$};

            \draw[-] (-.8,.2) -- (-.4,.2);
            \draw[<-] (.4,.2) -- (.8,.2) ;

            \draw[-] (-.8,.2) -- (-.8,.9);
            \draw[-] (.8,.2) -- (.8,.9);

            \draw[-] (.8,.9) -- (.3,.9);
            \draw[->] (-.8,.9) -- (-.3,.9);

            \node at (-1.1,.5) {$z_\mathrm{d}$};
            \node at (1.1,.5) {$w_\mathrm{d}$};
        \end{tikzpicture}
        \caption{A system in LFR form.}
        \label{fig:lfr}
    \end{subfigure} 
    \begin{subfigure}[b]{\linewidth}
        \vspace{1em}
        \centering

        \tikzstyle{block} = [draw, rectangle, 
        minimum height=1.5em, minimum width=2em]
        \tikzstyle{sum} = [draw, circle, radius=.15, node distance={0.5cm and 0.5cm}]
        \tikzstyle{input} = [coordinate]
        \tikzstyle{output} = [coordinate]
        \tikzstyle{pinstyle} = [pin edge={to-,thin,black}]

        \scalebox{1}{
        \begin{tikzpicture}[auto, node distance = {0.2cm and 0.5cm}]
            \node [input, name=input] {};
            \node [coordinate, right = of input, name=input_junction] {};
            \node [block, right = of input_junction] (Pzr) {$P_\mathrm{dw}$};
            \node [sum, minimum size=0.3cm, inner sep=0pt, right = of Pzr] (sum) {$+$};
            \node [block, right = of sum] (lti) {$\Phi$};
            \node [block, above = of lti] (Per) {$P_\mathrm{zw}$};
            \node [coordinate, right = of lti] (z_intersection) {};
            \node [block, right = of z_intersection] (Pew) {$P_\mathrm{zd}$};
            \node [sum, minimum size=0.3cm, inner sep=0pt, right = of Pew] (sum_e) {$+$};
            \node [output, right = of z_intersection] (output) {}; %
            \node [block, below = of lti] (static_nl) {$P_\mathrm{dd}$};
            \node [output, right = of sum_e] (output_e) {};
        
            \draw [->] (input) -- node {$ $} (Pzr);
            \draw [-] (input) -- node {$w$} (input_junction);
            \draw [->] (Pzr) -- node {$ $} (sum);
            \draw [->] (input_junction) |- (Per);
            \draw [->] (Per) -| (sum_e);
            \draw [->] (Pew) -- node[pos=0.8] {$ $} (sum_e);
            \draw [->] (sum) -- node {$z_\mathrm{d}$} (lti);
            \draw [->] (lti) -- node [name=z] {$w_\mathrm{d}$} (output);
            \draw [->] (z) |- (static_nl);
            \draw [->] (static_nl) -| node[pos=0.95] {$ $} (sum);
            \draw [->] (sum_e) -- node {$z$} (output_e);
        \end{tikzpicture}
        }
        \caption{Block diagram representation with individual channels.}
        \label{fig:lfr_block}
    \end{subfigure}
    \caption{The LFR interconnection.}
    \label{fig:LFRs}
    \vspace{-0.25cm}
\end{figure}
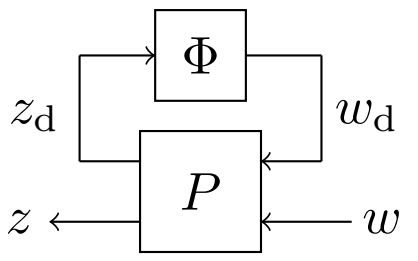
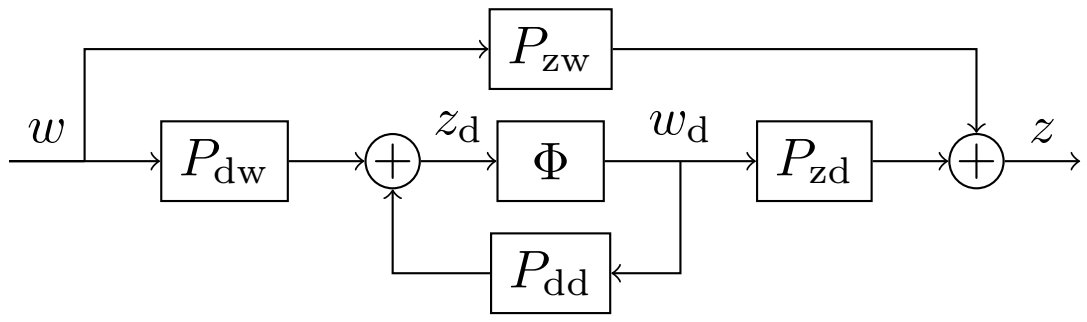

We are interested in systems in \emph{Linear Fractional Representation} (LFR) form, which is depicted in Fig.~\ref{fig:LFRs}. In an LFR, $\Phi : \Lte^k \to \Lte^m$ is a \emph{stable and causal} nonlinear and/or uncertain operator, and $P \in RH_\infty^{(k+q) \times (m+p)}$ is defined as $\mat{z_\mathrm{d} \\ z} = \mat{P_\mathrm{dd} & P_\mathrm{dw} \\ P_\mathrm{zd} & P_\mathrm{zw} } \mat{w_\mathrm{d} \\ w}$, where, for all $t \in \R_{\geq 0}$, $w(t) \in \R^p$ is the disturbance channel, $z(t) \in \R^q$ is the performance channel, and $z_\mathrm{d}(t) \in \R^k$ and $w_\mathrm{d}(t) \in \R^m$ are the input and output of $\Phi$, respectively. Let $\Phi \star P$ denote the interconnection $w_\mathrm{d} = \Phi z_\mathrm{d}$, then the closed-loop system
\begin{equation}\label{eq:lfr_closed_loop}
    \Phi \star P = P_\mathrm{zd} \Phi (I - P_\mathrm{dd} \Phi)^{-1} P_\mathrm{dw} + P_\mathrm{zw}
\end{equation}
is said to be \emph{well-posed} if $I - P_\mathrm{dd} \Phi$ has a causal inverse on $\Lte^k$~\cite{megretskiSystemAnalysisIntegral1997}, and is said to be \emph{stable} if the map $w \mapsto (z,z_\mathrm{d},w_\mathrm{d})$ has finite $L_2$-gain.

\subsection{Integral Quadratic Constraints}\label{sec:iqc_analysis}

Let $\Delta : \mathcal{H}_1 \to \mathcal{H}_2$ be an operator between Hilbert spaces $\mathcal{H}_1,\mathcal{H}_2$, and let $\inner{\cdot}{\cdot}_\mathcal{H} : \mathcal{H} \times \mathcal{H} \to \C$ be the inner product of the product Hilbert space $\mathcal{H} = \mathcal{H}_1 \times \mathcal{H}_2$, and $\Pi$ be a bounded linear self-adjoint operator on $\mathcal{H}$. Then, $\Delta$ is said to satisfy the IQC defined by the multiplier $\Pi$ if~\cite{jonssonLectureNotesIntegral2001}
\begin{equation}\label{eq:iqc_def}
    \sigma_\Pi(\Delta(v),v) := \inner{\mat{v \\ \Delta(v)}}{\Pi \mat{v \\ \Delta(v)}} \geq 0,
\end{equation}
for all $v \in \mathcal{H}_1$. For a given multiplier $\Pi$, the set of operators $\Delta$ that satisfy its IQC is denoted as $\mathrm{IQC}(\Pi)$. Since IQCs describe possibly nonlinear and/or uncertain dynamics using quadratic constraints, they allow for convex synthesis tools~\cite{megretskiSystemAnalysisIntegral1997,veenmanRobustStabilityPerformance2016,schererDissipativityIntegralQuadratic2022}. At the same time, as a multiplier describes a \emph{set} of operators, instead of only \emph{one} specific operator, IQC analysis is inherently conservative/robust. Conservatism can be reduced by finding a family of multipliers $\mathbf{\Pi}$, reducing the set of described operators to $\cap_{\Pi \in \mathbf{\Pi}} \mathrm{IQC}(\Pi)$.

If $\mathcal{H} = L_2^k \times L_2^m$, the IQC in~\eqref{eq:iqc_def} can be expressed in the Fourier domain via the Plancherel theorem as
\begin{equation}\label{eq:iqc_fourier}
    \int_\R \mat{\hat{v}(j \omega) \\ \widehat{\Delta(v)}(j \omega)}^* \Pi(j \omega) \mat{\hat{v}(j \omega) \\ \widehat{\Delta(v)}(j \omega)} \dd \omega \geq 0, 
\end{equation} 
for all $v \in L_2^k$, where $\hat{v}(j \omega) = \int_0^\infty v(\tau) e^{-j \omega \tau} \dd \tau$ with $\omega \in \R$ is the Fourier transform on $L_2$. %
The frequency-domain representation in~\eqref{eq:iqc_fourier} is used in~\cite{megretskiSystemAnalysisIntegral1997}, where $\Pi : j \R \to \C^{(k+m)\times(k+m)}$ is a bounded Hermitian-valued function. The advantage of this representation is that one can pick, frequency-wise, any $\Pi(j \omega)$ to describe a desired frequency-domain behavior, which is useful for engineering purposes~\cite{jonssonAnalysisPeriodicallyForced2003}. However, for computational purposes, we restrict ourselves to multipliers that can be represented using finite-dimensional dynamical systems $\Pi = \Psi^* M \Psi$, where $\Psi \in RH_\infty^{v \times (k+m)}$ and $M \in \mathbb{S}^v$. See~\cite{veenmanRobustStabilityPerformance2016} for details on how to obtain multipliers for nonlinearities and/or uncertainties.

One can also describe the \emph{desired} behavior using a performance multiplier~\cite{veenmanRobustStabilityPerformance2016}. In this paper, we focus on \emph{Quadratic Supply Rate} (QSR) performance with multipliers $\Pi_\mathrm{QSR}$
\begin{equation}\label{eq:qsr_performance}
    \sigma_{\Pi_\mathrm{QSR}}(w,z) \geq 0, \quad \Pi_\mathrm{QSR} := \mat{Q & S \\ S^\top & R},
\end{equation}
for all $w \in L_2^p$ with $z = (\Phi \star P)w$ in~\eqref{eq:lfr_closed_loop}, where $Q\in \mathbb{S}^q, S \in \R^{q \times p}$ and $R \in \mathbb{S}^p$. To bound the $L_2$-gain in~\eqref{eq:L2-gain}, define $\Pi_\gamma$ for $\gamma \geq 0$ as $\Pi_\mathrm{QSR}$ with $Q = -I,S=0$, and $R = \gamma^2 I$. Then, $\sigma_{\Pi_\gamma}(w,z) \geq 0 \iff \norm{z} \leq \gamma \norm{w}$.

The following proposition, see~\cite[Cor. 3]{veenmanRobustStabilityPerformance2016},~\cite[Eq. (40)]{schererDissipativityIntegralQuadratic2022}, demonstrates how IQC analysis can be used to verify stability and QSR performance of~\eqref{eq:lfr_closed_loop}.

\begin{proposition}\label{prop:L2_performance_iqc}
    Consider the LFR in~\eqref{eq:lfr_closed_loop} with $P \in RH_\infty^{(k+q) \times (m+p)}$, bounded and causal $\Phi : L_2^k \to L_2^m$, bounded self-adjoint operators $\Pi_l$ on $L_2^{k+m}$ for $l=1,\dots,N$, and $0 \geq Q\in \mathbb{S}^q, S \in \R^{q \times p}$ and $R \in \mathbb{S}^p$. If 
    \begin{enumerate}
        \item The LFR $(\tau \Phi) \star P$ is well-posed for $\tau \in [0,1]$,
        \item For all $l=1,\dots ,N$ and $\tau \in [0,1]$, the map $\tau \Phi$ satisfies the IQC $\sigma_{\Pi_l}(\tau \Phi(z),z) \geq 0$ for all $z \in L_2^k$,
        \item There exist $\tau_l \geq 0$ for $l=1,\dots,N$ such that the following FDI holds for all $\omega \in \R \cup \{ \infty \}$: 
    \end{enumerate}
    \begin{equation}\label{eq:iqc_fdi}
        \scalebox{.84}{$(\star)^* \mat{\sum^N_{l=1} \tau_l \Pi_l(j \omega) & 0 \\ 0& -\Pi_\mathrm{QSR}  } \mat{P_\mathrm{dd}(j \omega) & P_\mathrm{dw}(j \omega) \\ I & 0 \\ P_\mathrm{zd}(j \omega) & P_\mathrm{zw}(j \omega) \\ 0 & I} < 0$}.
    \end{equation}
    Then,~\eqref{eq:lfr_closed_loop} is stable and has QSR performance.
\end{proposition}

Analogous to~\eqref{eq:iqc_def}, an operator $\Delta : \mathcal{H}_1 \to \mathcal{H}_2$ is said to satisfy the incremental IQC (IIQC) for a multiplier $\Pi$ if 
\begin{equation}\label{eq:iiqc_def}
    \inner{\mat{v_1-v_2 \\ \Delta(v_1) - \Delta(v_2)}}{\Pi \mat{v_1-v_2 \\ \Delta(v_1) - \Delta(v_2)}} \geq 0,
\end{equation}
for all $v_1,v_2 \in \mathcal{H}_1$~\cite[Def. 3]{jonssonAnalysisPeriodicallyForced2003}. Then, Proposition~\ref{prop:L2_performance_iqc} with IIQCs $\Pi_l$ for $\Phi$ and the performance IIQC $\Pi_\gamma$ yields an incremental $L_2$-gain bound $\Gamma(\Phi \star P) \leq \gamma$.

\subsection{The Kalman-Yakubovich-Popov Lemma}

The condition in~\eqref{eq:iqc_fdi} amounts to verifying uncountably infinitely many matrix inequalities. When the multipliers $\Pi_l$ have finite-dimensional representations, the KYP lemma, see~\cite[Thm. 1]{rantzerKalmanYakubovichPopov1996}, provides an equivalent finite-dimensional convex feasibility test in terms of an LMI.

\begin{lemma}\label{lemma:kyp}
    Let $A \in \R^{n \times n}, B \in \R^{n \times m}, M \in \mathbb{S}^{n+m}$ with $\det(j \omega I-A) \neq 0$ for $\omega \in \R$, then 
    \begin{equation}
        (\star)^* M \mat{(j \omega I - A)^{-1} B \\ I} < 0, \quad \forall \omega \in \R \cup \{ \infty \},
    \end{equation}
    is equivalent to $\exists P \in \mathbb{S}^n$ such that
    \begin{equation}
        M + (\star)^* \mat{0 & P \\ P & 0} \mat{A & B \\ I & 0} < 0.
    \end{equation}
\end{lemma}

The GKYP lemma~\cite[Thm. 2]{iwasakiGeneralizedKYPLemma2005} provides an equivalent result for certain frequency ranges in $\R$.

\begin{lemma}\label{lemma:gkyp}%
    Let $A \in \R^{n \times n}, B \in \R^{n \times m}, M \in \mathbb{S}^{n+m}$ with $(A,B)$ controllable and $\varpi >0$, then 
    \begin{equation}\label{eq:gkyp_fdi}
        (\star)^* M \mat{(j \omega I - A)^{-1} B \\ I} \leq 0, %
    \end{equation}
    for all $\omega \in [\varpi, \infty]$ such that $\det(j \omega I-A) \neq 0$ is equivalent to $\exists P, Q \in \mathbb{S}^n$ with $Q \geq 0$ such that
    \begin{equation}\label{eq:gkyp_lmi}
        M + (\star)^* \mat{Q & P \\ P & - \varpi^2 Q} \mat{A & B \\ I & 0} \leq 0.
    \end{equation}
    Similarly,~\eqref{eq:gkyp_fdi} for $\omega \in [0,\varpi]$ such that $\det(j \omega I-A) \neq 0$ is equivalent to $\exists P, Q \in \mathbb{S}^n$ with $Q \geq 0$ such that
    \begin{equation}\label{eq:gkyp_lmi_low}
        M + (\star)^* \mat{-Q & P \\ P &  \varpi^2 Q} \mat{A & B \\ I & 0} \leq 0.
    \end{equation}
\end{lemma}

\section{The Nonlinear Bode Diagram}\label{sec:gammas_bode_def}

In this section, we will focus on a class of systems that are suitable for frequency-domain performance analysis, which are systems that preserve the periodicity of the input. For these systems, we define frequency-dependent $L_2$-gains, which form the basis of the nonlinear Bode diagram.

\subsection{Period-Preserving Systems}

Operators that yield an asymptotically $T$-periodic output, given a $T$-periodic input, are called \emph{period-preserving}.

\begin{definition}\label{def:period-preserving}
    An operator $R : \Lte^p \to \Lte^q$ is called period-preserving if, for every $T$-periodic input $w \in \Lte^p$, the output $z=Rw$ converges asymptotically to a $T$-periodic signal $\tilde{z} : \R_{\geq 0} \to \R^q$, i.e., $\forall \epsilon>0 , \exists t>0 : \forall \tau \geq t, \; |z(\tau)-\tilde{z}(\tau)| < \epsilon$.
\end{definition}

The LFR in~\eqref{eq:lfr_closed_loop} is called period-preserving if $w \mapsto (z,z_\mathrm{d},w_\mathrm{d})$ is period-preserving. 

Examples of systems that have the period-preserving property are convergent systems~\cite{pavlovUniformOutputRegulation2006}, systems with fading memory~\cite{boydFadingMemoryProblem1985}, or (strongly) contractive systems~\cite{lohmillerContractionAnalysisNonlinear1998,bulloContractionTheoryDynamical2023}. The incremental IQC conditions in~\cite[Prop. 2]{jonssonAnalysisPeriodicallyForced2003} require additional smoothness assumptions to guarantee pointwise asymptotic convergence to a periodic solution.

Traditionally, a Bode diagram measures the gain from periodic input $w$ to steady-state output $\tilde{z}$~\cite{skogestadMultivariableFeedbackControl2010}. To define a nonlinear Bode diagram with a similar interpretation, we need a relationship between the steady-state $\tilde{z}$, and the limit $t \to \infty$ for $z=Rw$, which includes transient behavior.

\begin{lemma}\label{lemma:pp-lemma}
    Let $R : \Lte^p \to \Lte^q$ be a causal and period-preserving system with $\gamma(R) <\infty$. For any $T$-periodic $0 \neq w \in \Lte^p$, let $\tilde{z}$ denote the $T$-periodic asymptote of $z=Rw$. Then, %
    \begin{equation}\label{eq:pp_lemma_z_ztilde_eq}
        \maketextstyle \lim_{t \to \infty} \frac{\norm{P_t z}}{\norm{P_t w}} = \frac{\norm{\tilde{z}}_\mathrm{RMS}}{\norm{w}_\mathrm{RMS}} \leq \gamma(R) , \quad t \in \R_{\geq 0}.
    \end{equation}
\end{lemma}

\begin{proof}
    Define $\norm{x}_{[t_1,t_2]} := \sqrt{\int_{t_1}^{t_2} x(\tau)^\top x(\tau) \dd \tau}$, and since $\sqrt{a^2+b^2} \leq a+b$ for $a,b \in \R_{\geq 0}$, we have the inequality 
    \begin{multline*}
        \textstyle \norm{w}_{[0,t_1+t_2]} = \sqrt{\int_0^{t_1} |w(\tau)|^2 \dd \tau + \int_{t_1}^{t_1+t_2} |w(\tau)|^2 \dd \tau} \\ \leq  \norm{w}_{[0,t_1]} + \norm{w}_{[t_1,t_1+t_2]}.
    \end{multline*}

    First, we prove $\tilde{z} \in \Lte^q$. Fix some $\epsilon >0$. Since $R$ is period-preserving, $\norm{\tilde{z}-z}_{[t_\epsilon,t_\epsilon+T]}  \leq \epsilon \sqrt{T}$ for some $t_\epsilon \in \R_{\geq 0}$. Since $\gamma(R)<\infty$ and $R$ causal, it holds that $z = Rw \in \Lte^q$, hence $\norm{z}_{[t_\epsilon,t_\epsilon+T]} < \infty$. As both $z$ and $\tilde{z} - z$ for $t \in [t_\epsilon,t_\epsilon+T]$ are members of the vector space $L_2^q[0,T]$ after translation by $t_\epsilon$, so is $\tilde{z} = z + (\tilde{z}-z)$, hence $\norm{\tilde{z}}_{[0,T]} < \infty$, proving $\tilde{z} \in \Lte^q$.%

    Then, we prove the equality in~\eqref{eq:pp_lemma_z_ztilde_eq} by showing the inequality holds both ways. For the remainder of the proof, let $t_\tau := (\tau+\tau^2)T$ for $\tau \in \R_{\geq 0}$, and $\overline{\tau} := \lceil \tau \rceil, \underline{\tau} := \lfloor \tau \rfloor$. Let $e(t) := |z(t)-\tilde{z}(t)|$ and define $\tilde{\epsilon}(t) : = \sup_{s \geq t} e(s)$, then $\tilde{\epsilon}(t) \xrightarrow{t \to \infty} 0$ by the asymptotic convergence of $z \to \tilde{z}$. To prove $\lim_{t \to \infty} \frac{\norm{P_t z}}{\norm{P_t w}} \leq \frac{\norm{\tilde{z}}_\mathrm{RMS}}{\norm{w}_\mathrm{RMS}}$, consider
    \begin{multline*}
        \norm{z}_{[0,t_\tau]} \leq \norm{z}_{[0,\overline{\tau}T]} +  \norm{\tilde{z}}_{[\underline{\tau} T, t_{\overline{\tau}}]} + \norm{z-\tilde{z}}_{[\underline{\tau} T, t_{\overline{\tau}}]} \\
        \leq \sqrt{\overline{\tau}} \gamma(R) \norm{w}_{[0,T]} + \sqrt{1+\overline{\tau}^2 }(\norm{\tilde{z}}_{[0,T]} + \tilde{\epsilon}(\underline{\tau} T) \sqrt{T}),
    \end{multline*}
    then for $t > 2T$\footnote{To avoid $\underline{\tau} = 0$, we need $\tau >1$, which is equivalent to $t > 2T$.}, one has
    \begin{multline*}
        \textstyle \lim_{t \to \infty} \frac{\norm{P_{t} z}}{\norm{P_{t} w}} = \lim_{\tau \to \infty} \frac{\norm{z}_{[0,t_\tau]}}{\norm{w}_{[0,t_\tau]}} \leq \lim_{\tau \to \infty} \frac{\norm{z}_{[0,t_\tau]}}{\norm{w}_{[0,t_{\underline{\tau}}]}}  \\
        \textstyle \leq \lim_{\tau \to \infty } \frac{\sqrt{\overline{\tau}} \gamma(R) \norm{w}_{[0,T]} + \sqrt{1+\overline{\tau}^2 }(\norm{\tilde{z}}_{[0,T]} + \tilde{\epsilon}(\underline{\tau} T) \sqrt{T})}{\sqrt{\underline{\tau}+\underline{\tau}^2} \norm{w}_{[0,T]}}  \\ 
        \textstyle = \frac{\norm{\tilde{z}}_{[0,T]}}{\norm{w}_{[0,T]}} = \frac{\norm{\tilde{z}}_\mathrm{RMS}}{\norm{w}_\mathrm{RMS}},
    \end{multline*}
    as $\frac{\sqrt{\overline{\tau}}}{\sqrt{\underline{\tau}+\underline{\tau}^2} } \xrightarrow{\tau \to \infty} 0, \frac{\sqrt{1 + \overline{\tau}^2}}{\sqrt{\underline{\tau}+\underline{\tau}^2}} \xrightarrow{\tau \to \infty} 1$, and $ \tilde{\epsilon}(\underline{\tau} T)\xrightarrow{\tau \to \infty} 0$. To prove $\frac{\norm{\tilde{z}}_\mathrm{RMS}}{\norm{w}_\mathrm{RMS}} \leq \lim_{t \to \infty} \frac{\norm{P_t z}}{\norm{P_t w}} \leq \gamma(R)$, observe that
    \begin{multline*}
        \norm{\tilde{z}}_{[0,t_\tau]} \leq \norm{z}_{[0,t_\tau]} + \norm{z-\tilde{z}}_{[0,\overline{\tau}T]} + \norm{z-\tilde{z}}_{[\underline{\tau}T,t_{\overline{\tau}}]}  \\
        \leq \norm{z}_{[0,t_\tau]} + \norm{z}_{[0,\overline{\tau}T]} + \norm{\tilde{z}}_{[0,\overline{\tau}T]}+ \norm{z-\tilde{z}}_{[\underline{\tau}T,t_{\overline{\tau}}]}  \\
        \textstyle \leq \norm{z}_{[0,t_\tau]} + \sqrt{\overline{\tau}}(\gamma(R)\norm{w}_{[0,T]} + \norm{\tilde{z}}_{[0,T]}) \\ \textstyle + \sqrt{1+\overline{\tau}^2} \tilde{\epsilon}(\underline{\tau}T)  \sqrt{T}.
    \end{multline*}
    For $t >2T$, the latter yields
    \begin{multline*}
        \textstyle \frac{\norm{\tilde{z}}_\mathrm{RMS}}{\norm{w}_\mathrm{RMS}} = \sqrt{\frac{\overline{\tau} + \overline{\tau}^2}{\underline{\tau} + \underline{\tau}^2}}  \frac{\norm{\tilde{z}}_{[0,t_{\underline{\tau}}]}}{\norm{w}_{[0,t_{\overline{\tau}}]}}  \leq c(\tau) \frac{\norm{\tilde{z}}_{[0,t_\tau]}}{\norm{w}_{[0,t_{\tau}]}}  \\ 
        \textstyle \leq  c(\tau) \frac{\norm{z}_{[0,t_\tau]} + \sqrt{\overline{\tau}}(\gamma(R)\norm{w}_{[0,T]} + \norm{\tilde{z}}_{[0,T]})  + \sqrt{1+\overline{\tau}^2} \tilde{\epsilon}(\underline{\tau}T)  \sqrt{T}}{\norm{w}_{[0,t_{\tau}]}} \\
        \textstyle \leq  c(\tau) \left[ \frac{\norm{z}_{[0,t_\tau]} }{\norm{w}_{[0,t_{\tau}]}} + \frac{ \sqrt{\overline{\tau}}(\gamma(R)\norm{w}_{[0,T]} + \norm{\tilde{z}}_{[0,T]})  + \sqrt{1+\overline{\tau}^2} \tilde{\epsilon}(\underline{\tau}T)  \sqrt{T}}{\sqrt{\underline{\tau} + \underline{\tau}^2} \norm{w}_{[0,T]}} \right] \\
        \textstyle \xrightarrow{\tau \to \infty} \lim_{\tau \to \infty} \frac{\norm{z}_{[0,t_{\tau}]}}{\norm{w}_{[0,t_{\tau}]}} = \lim_{t \to \infty} \frac{\norm{P_t z}}{\norm{P_t w}},
    \end{multline*}
    where $c(\tau) := \sqrt{\frac{\overline{\tau} + \overline{\tau}^2}{\underline{\tau} + \underline{\tau}^2}}  \xrightarrow{\tau \to \infty} 1$ and $\norm{\tilde{z}}_{[0,T]} < \infty$ is used, proving $\lim_{t \to \infty} \frac{\norm{P_t z}}{\norm{P_t w}} = \frac{\norm{\tilde{z}}_\mathrm{RMS}}{\norm{w}_\mathrm{RMS}}$. Finally,~\eqref{eq:pp_lemma_z_ztilde_eq} follows from $\norm{P_t z} = \norm{P_t R P_t w} \leq \norm{R P_t w} \leq \gamma(R) \norm{P_t w}$.
\end{proof}

Lemma~\ref{lemma:pp-lemma} establishes two important facts: First, the periodic steady-state $\tilde{z} : \R_{\geq 0} \to \R^q$ is a member of $\Lte^q$. Second, the equivalence between the time-domain limit and the RMS-gain in~\eqref{eq:pp_lemma_z_ztilde_eq} bridges the time-domain performance, which includes transients, and the frequency-domain RMS performance. 

A similar result is given in~\cite[Lemma 1]{krebbekxNonlinearBandwidthBode2025a} for the incremental gain, however, Lemma~\ref{lemma:pp-lemma} adds two important results. First, $\tilde{z} \in \Lte^q$ is proven, which is \emph{assumed} in~\cite{krebbekxNonlinearBandwidthBode2025a}. Second, the limit in~\eqref{eq:pp_lemma_z_ztilde_eq} is proven for $t \in \R_{\geq 0}$, while in~\cite{krebbekxNonlinearBandwidthBode2025a} it is only shown for $t=nT$ with $n \in \N_+$ as $n \to \infty$. The latter is important, since it can otherwise occur that different subsequences for $t \in \R_{\geq 0}$ in~\eqref{eq:pp_lemma_z_ztilde_eq} converge to different limits~\cite{mariCounterexamplePowerSignals1996}.

Note that Lemma~\ref{lemma:pp-lemma} does not depend on the method that guarantees the period-preserving property. Therefore, it can be used to bridge the steady-state and time-domain performance for the methods in~\cite{pavlovUniformOutputRegulation2006,lohmillerContractionAnalysisNonlinear1998,bulloContractionTheoryDynamical2023,jonssonAnalysisPeriodicallyForced2003,boydFadingMemoryProblem1985}.

\subsection{Frequency-Dependent Gain}

For periodic signals, Lemma~\ref{lemma:pp-lemma} shows that the frequency-domain performance can be quantified using the RMS-gain $\frac{\norm{\tilde{z}}_\mathrm{RMS}}{\norm{w}_\mathrm{RMS}}$. However,~\eqref{eq:pp_lemma_z_ztilde_eq} merely provides the bound $\gamma(R)$, which holds for all periods of the input. Instead, we introduce the following sets of input signals, analogous to~\cite{krebbekxNonlinearBandwidthBode2025a}.

\begin{definition}
    For a frequency $\omega > 0$, we define
    \begin{subequations}\label{eqs:input_spaces}
    \begin{align}
        \mathcal{U}_{\omega \mathrm{e}}^d &:= \maketextstyle  \{ u\in \Lte^d \mid u(t) =  \hat{u}_1 e^{j \omega t} + \bar{\hat{u}}_1 e^{-j\omega t} \}, \\
        \overline{\mathcal{U}}^d_{\omega \mathrm{e}} &:= \maketextstyle \{ u\in \Lte^d \mid u(t) = \sum_{0 \neq n \in \Z} \hat u_n e^{j \omega n t} \}, \\
        \begin{split}\label{eq:U_subharmonic}
        \underline{\mathcal{U}}^d_{\omega \mathrm{e}} &:= \maketextstyle \{  u\in \Lte^d \mid \\  & \maketextstyle u(t) = \sum_{n=-M,n \neq 0}^M \hat u_n e^{j (\omega  / M) n t}, \, M \in \N_{+}  \},
        \end{split}
    \end{align}
    \end{subequations}
    which are called the sinusoidal, harmonic and subharmonic input spaces, respectively, $\hat{u}_{-n}=\bar{\hat{u}}_{n} \in \C^d$ are the Fourier coefficients, and $T = 2 \pi / \omega$ is the period. 
\end{definition}

We define the Hilbert space $\overline{\mathcal{U}}^d_\omega := P_T \overline{\mathcal{U}}^d_{\omega \mathrm{e}} \subseteq L_2^d[0,T]$, where each $u \in \overline{\mathcal{U}}^d_{\omega \mathrm{e}}$ is uniquely identified with $P_T u = v \in \overline{\mathcal{U}}^d_{\omega}$, and vice versa by $u = \sum_{k =0}^\infty v(t-kT)$. The Hilbert spaces $\mathcal{U}_\omega^d$ and $\underline{\mathcal{U}}_\omega^d$ are similarly defined.

The requirement $M \in \N_+$ in~\eqref{eq:U_subharmonic}, which differs from~\cite{krebbekxNonlinearBandwidthBode2025a}, ensures that $w \in \underline{\mathcal{U}}^d_{\omega \mathrm{e}}$ has period $T = 2 \pi M / \omega $. If $M$ were allowed to be infinite, then $w$ is not guaranteed to be periodic, making both $\norm{w}_\mathrm{RMS}$ and the steady-state $\tilde{z}$ ill-defined.

By restricting the input in~\eqref{eq:pp_lemma_z_ztilde_eq} to the spaces in~\eqref{eqs:input_spaces}, we obtain the corresponding frequency-dependent gains.

\begin{definition}
    For a period-preserving operator $R : \Lte^p \to \Lte^q$, the frequency-dependent $L_2$-gains are defined as
    \begin{subequations}\label{eqs:gammas}
    \begin{align}
        \gamma_\omega(R) &= \maketextstyle \sup_{0\neq w \in \mathcal{U}_{\omega \mathrm{e}}^p} \frac{\norm{\tilde{z}}_\mathrm{RMS}}{\norm{w}_\mathrm{RMS}}, \label{eq:gamma_sinus} \\
        \overline{\gamma}_\omega(R) &= \maketextstyle \sup_{0 \neq w \in \overline{\mathcal{U}}^p_{\omega \mathrm{e}}} \frac{\norm{\tilde{z}}_\mathrm{RMS}}{\norm{w}_\mathrm{RMS}}, \label{eq:gamma_harmonic} \\
        \textstyle \underline{\gamma}_\omega(R) &= \maketextstyle \sup_{0 \neq w \in \underline{\mathcal{U}}^p_{\omega \mathrm{e}}} \frac{\norm{\tilde{z}}_\mathrm{RMS}}{\norm{w}_\mathrm{RMS}}, \label{eq:gamma_subharmonic}
    \end{align}
    \end{subequations}
    where $\tilde{z}$ is the periodic asymptote of $z=Rw$.
\end{definition}

Since we have the inclusions $\mathcal{U}_{\omega \mathrm{e}}^d \subseteq \overline{\mathcal{U}}_{\omega \mathrm{e}}^d \subseteq \Lte^d$ and $\mathcal{U}_{\omega \mathrm{e}}^d \subseteq \underline{\mathcal{U}}_{\omega \mathrm{e}}^d \subseteq \Lte^d$ for any $\omega \in \R_{>0}$, it follows that 
\begin{equation}\label{eq:sinus_harmonic_gain_inequalities}
    \gamma_\omega(R) \leq \overline{\gamma}_\omega(R) \leq \gamma(R), \, \gamma_\omega(R) \leq \underline{\gamma}_\omega(R) \leq \gamma(R).
\end{equation}
From~\eqref{eq:sinus_harmonic_gain_inequalities}, it is clear that the gains in~\eqref{eqs:gammas} provide a frequency-dependent refinement of the $L_2$-gain for nonlinear systems. The sinusoidal gain~\eqref{eq:gamma_sinus} can be seen as a non-approximative counterpart to the describing function from~\cite{krylovIntroductionNonlinearMechanics1943}, and reduces to the LTI gain $|R(j \omega)|$ if $R$ is LTI, while the harmonic gain~\eqref{eq:gamma_harmonic} includes generalizations thereof to multisine inputs as presented in~\cite{rijlaarsdamComparativeOverviewFrequency2017}. Note, however, that the gains in~\eqref{eqs:gammas} lack any notion of \emph{phase}. The subharmonic gain~\eqref{eq:gamma_subharmonic} is a novel concept in the analysis of nonlinear systems, and is particularly useful for characterizing low-frequency behavior, such as integrator behavior, or a finite DC gain.

\subsection{The Bode Diagram for Nonlinear Systems}

Let $R : \Lte^p \to \Lte^q$ be a MIMO operator where $z = Rw$, define the SISO map $R_{j \leftarrow i} : \Lte \to \Lte$ as 
\begin{equation}\label{eq:siso_transfers}
    R_{j \leftarrow i} = (\underbrace{0 \cdots 0}_{j-1} \ 1 \ 0 \cdots 0)R (\underbrace{0 \cdots 0}_{i-1} \ 1 \ 0 \ \cdots 0)^\top,
\end{equation}
which is $R$ where only input $i \in \{1,\dots, p\}$ and output $j \in \{1,\dots,q\}$ is selected. While an LTI system is fully characterized by the transfers in~\eqref{eq:siso_transfers} since $z_j = \sum_{i=1}^p R_{j \leftarrow i} w_i$ by linearity, this is not always true for nonlinear systems: e.g., $R:(x(t),y(t)) \mapsto x(t)y(t)$. 

To capture the effect of multiple nonzero inputs on the output, let $R_j : w \mapsto z_j$ denote the map from all inputs $w$ to the output $z_j$. Then, we use an input filter $W_\mathrm{in} \in RH_\infty^{p \times p}$ chosen such that all expected input signals $w$ are obtained as $w = W_\mathrm{in} w_\mathrm{n}$, where $\norm{w_\mathrm{n}}_\mathrm{RMS} \leq 1$ is normalized. This is a standard approach in the robust control framework, see~\cite[Ch. 9.3.6]{skogestadMultivariableFeedbackControl2010} or~\cite[Ch. 5.7]{zhouRobustOptimalControl1996}. Using this filter, the gains in~\eqref{eqs:gammas} for $R_j W_\mathrm{in} : w_\mathrm{n} \mapsto z_j$ provide a frequency-dependent upper bound on the expected norm of $z_j$.

\begin{definition}\label{def:bode}
    The nonlinear Bode diagram for a causal and stable $R : \Lte^p \to \Lte^q$ is the matrix formed by plots of the gains in~\eqref{eqs:gammas} for the SISO maps $R_{j \leftarrow i}$ in~\eqref{eq:siso_transfers}, where $j$ and $i$ are the row and column indices, respectively, with an additional column formed by plotting~\eqref{eqs:gammas} for $R_j W_\mathrm{in}$.
\end{definition}

While $W_\mathrm{in}$ captures the expected frequency content of the inputs, an output filter $W_\mathrm{out} \in RH_\infty^{q \times q}$ is used in the robust control framework to describe the desired performance. The performance objectives are met if $\norm{W_\mathrm{out} R W_\mathrm{in}}_\infty \leq 1$ holds.

The output filter for the $j$th output channel $z_j$, denoted as $W_{\mathrm{out},j}$, may be used in the nonlinear Bode diagram in the following way. In the Bode diagram of $R_j W_\mathrm{in}$, one can plot $|W_{\mathrm{out},j}(j \omega)|^{-1}$, as the inverse of the output filter corresponds to the allowed frequency content of the output. Then, one can verify if the sinusoidal gain satisfies $\gamma_\omega(R_j W_\mathrm{in}) \leq |W_{\mathrm{out},j}(j \omega)|^{-1}$ at all frequencies $\omega \in \R_+$. If this inequality holds, it means that the power gain from input $w$ to the first harmonic of the output is bounded by the weighting filter: $\sqrt{|\hat{z}_{j,1}|^2 + |\hat{z}_{j,-1}|^2} \leq |W_{\mathrm{out},j}(j \omega)|^{-1} \sqrt{|\hat{w}_1|^2 + |\hat{w}_{-1}|^2}$. Note, however, that since $R$ is not LTI per se, higher harmonics in the output may result in $\gamma(W_\mathrm{out} R W_\mathrm{in}) >1$. In that case, one can plot the Bode diagram of $(W_{\mathrm{out}} R W_\mathrm{in})_j$, which is the weighted closed-loop plant with only output channel $j$, to see at what frequency the gain exceeds one.

The inverse output filter profiles $|W_{\mathrm{out},j}(j \omega)|^{-1} $ are typically monotone, either increasing or decreasing. When it is increasing, it can be compared against the subharmonic gain $\underline{\gamma}_\omega(R_j W_\mathrm{in})$, as this is also a monotone increasing function in $\omega$. Conversely, a decreasing profile should be compared against the harmonic gain.

A Bode diagram similar to Definition~\ref{def:bode} was introduced in~\cite{krebbekxNonlinearBandwidthBode2025a}. However, the definition in~\cite{krebbekxNonlinearBandwidthBode2025a} was only valid for SISO systems with one nonlinearity, and used a different definition of the subharmonic input space. Moreover, the computation in~\cite{krebbekxNonlinearBandwidthBode2025a} was based on scaled relative graphs (see~\cite{ryuScaledRelativeGraphs2022,chaffeyGraphicalNonlinearSystem2023}), which require numerical complex geometry calculations. In contrast, in this paper we compute the nonlinear Bode diagram via the IQC framework, which yields a computationally tractable toolchain based on semidefinite programming.

\section{Computation of the Nonlinear Bode Diagram}\label{sec:nl_bode_computation}

In this section, we show how to compute upper bounds for the frequency-dependent gains in~\eqref{eqs:gammas} for systems in LFR form, using IQCs and the GKYP lemma. 

Throughout, we assume that~\eqref{eq:lfr_closed_loop} is well-posed, stable, and period-preserving. This guarantees, via Lemma~\ref{lemma:pp-lemma}, that for every $\omega>0$ and $w$ in~\eqref{eqs:input_spaces}, there exist unique periodic steady-state responses %
$(\tilde{z}^\top \ \tilde{z}_\mathrm{d}^\top \ \tilde{w}_\mathrm{d}^\top)^\top \in \Lte^{q+k+m}$.

\subsection{Computing the Harmonic Gain}\label{sec:harmonic_gain}

The map from $T$-periodic input $w \in \overline{\mathcal{U}}^p_{\omega \mathrm{e}}$ to $T$-periodic steady-state response $\tilde{z} \in \Lte^q$ can be viewed as $\tilde{\Phi} \star \tilde{P} : \overline{\mathcal{U}}^p_\omega \to L_2^q[0,T]$, where $T=\frac{2 \pi}{\omega}$ is the period and $\tilde{\Phi} : L_2^k[0,T] \to L_2^m[0,T]$ defines $\tilde{w}_\mathrm{d} = \tilde{\Phi} \tilde{z}_\mathrm{d}$, and $\tilde{P} : \sum_{n \in \Z} \hat{u}_n e^{ j \omega n t} \mapsto \sum_{n \in \Z} P(j \omega n) \hat{u}_n e^{ j \omega n t}$ maps $(\tilde{w}_\mathrm{d}^\top \ w^\top)^\top \mapsto (\tilde{z}_\mathrm{d}^\top \ \tilde{z}^\top)^\top$, where $\hat{u}_n = (\hat{\tilde{w}}_{\mathrm{d},n}^\top \ \hat{w}_n^\top)^\top$ and $P$ is from~\eqref{eq:lfr_closed_loop}. The IQC analysis of $\tilde{z} = \tilde{\Phi} \star \tilde{P} w$ is analogous to the one presented in Section~\ref{sec:iqc_analysis}, where the Hilbert space $L_2$ is now replaced by $L_2[0,T]$.

The steady-state map $\tilde{\Phi}$ is said to satisfy the IQC with bounded self-adjoint multiplier $\tilde{\Pi}$ on $L_2^{k+m}[0,T]$ if 
\begin{equation}\label{eq:iqc_harmonic}
    \sigma_{\tilde{\Pi}}^T(\tilde{\Phi}(z),z) := \int_0^T (\star)^\top \tilde{\Pi} \mat{z \\ \tilde{\Phi}(z)} \dd \tau \geq 0, \, \forall z \in L_2^k[0,T],
\end{equation}
which is equivalently represented in the Fourier domain as
\begin{equation}\label{eq:iqc_fourier_harmonic}
    \sum_{n \in \Z} (\star)^* \tilde{\Pi}(j \omega n) \mat{\hat{z}_n \\ \hat{w}_n} \geq 0, \, \forall z \in L_2^k[0,T], \, w = \tilde{\Phi} z.
\end{equation}
Moreover, $\tilde{\Phi} \star \tilde{P}$ is said to have QSR performance on $\overline{\mathcal{U}}_\omega^p$ if $\sigma^T_{\Pi_\mathrm{QSR}}(w,\tilde{z}) \geq 0$ for all $w \in \overline{\mathcal{U}}_\omega^p$.

It must be noted that if $\Phi : L_2^k \to L_2^m$ satisfies the IQC with multiplier $\Pi$, it is not always true that the corresponding steady-state map $\tilde{\Phi}$ satisfies the IQC with $\tilde{\Pi} = \Pi$, or vice versa~\cite{jonssonAnalysisPeriodicallyForced2003}. Equivalence does hold in certain cases, which includes the trivial case where $\Phi : \R^k \to \R^m$ is a static nonlinear function and $\Pi \in \mathbb{S}^{k+m}$ a static multiplier, and the nontrivial case where $\Phi$ is a static nonlinear function and $\Pi$ a ZF Multiplier~\cite[Ex. 3]{jonssonAnalysisPeriodicallyForced2003}.

\begin{theorem}\label{thm:harmonic_performance_iqc}
    Suppose that~\eqref{eq:lfr_closed_loop} is well-posed, stable, and period-preserving. For $\omega>0$, let $\tilde{\Pi}_l$ be bounded self-adjoint operators on $L_2^{k+m}[0,T]$ such that the map $\tilde{\Phi}$ satisfies the IQC $\sigma^T_{\tilde{\Pi}_l}(\tilde{\Phi}(z),z) \geq 0$ for all $z \in L_2^k[0,T]$ and $l=1,\dots,N$, and let $Q\in \mathbb{S}^q, S \in \R^{q \times p}$ and $R \in \mathbb{S}^p$.

    If there exist $\tau_l \geq 0$ for $l=1,\dots,N$ such that the following FDI holds for all $\tilde{\omega} = \omega n$, where $n \in \N_+$ 
    \begin{equation}\label{eq:iqc_fdi_harmonic}
        \scalebox{.84}{$(\star)^* \mat{\sum^N_{l=1} \tau_l \tilde{\Pi}_l(j \tilde{\omega}) & 0  \\ 0& -\Pi_\mathrm{QSR}  } \mat{P_\mathrm{dd}(j \tilde{\omega}) & P_\mathrm{dw}(j \tilde{\omega}) \\ I & 0 \\ P_\mathrm{zd}(j \tilde{\omega}) & P_\mathrm{zw}(j \tilde{\omega}) \\ 0 & I} \leq 0,$}
    \end{equation}
    and the following FDI holds at $\tilde{\omega} = 0$
    \begin{equation}\label{eq:iqc_fdi_zero_w}
        \scalebox{.84}{$(\star)^* \mat{\sum^N_{l=1} \tau_l \tilde{\Pi}_l(j \tilde{\omega}) & 0  \\ 0& -\Pi_\mathrm{QSR}  } \mat{P_\mathrm{dd}(j \tilde{\omega})   \\ I  \\ P_\mathrm{zd}(j \tilde{\omega})   \\ 0 } \leq 0,$}
    \end{equation}
    then $\tilde{\Phi} \star \tilde{P} : w \mapsto \tilde{z}$ has QSR performance for $w \in \overline{\mathcal{U}}_\omega^p$.
\end{theorem}

\begin{proof}
    As~\eqref{eq:lfr_closed_loop} is well-posed, stable, and period-preserving, we know via Lemma~\ref{lemma:pp-lemma} that $\forall w \in \overline{\mathcal{U}}_\omega^p$, there exist unique $T$-periodic steady-state signals $(\tilde{z}^\top \ \tilde{z}_\mathrm{d}^\top \ \tilde{w}_\mathrm{d}^\top)^\top \in L_2^{q+k+m}[0,T]$, where $T = \frac{2 \pi}{\omega}$ is the period, and the map $w \mapsto (\tilde{z}^\top \ \tilde{z}_\mathrm{d}^\top \ \tilde{w}_\mathrm{d}^\top)^\top$ has finite gain $\overline{\gamma}_\omega$. To verify QSR performance, note that~\eqref{eq:iqc_fdi_harmonic} and~\eqref{eq:iqc_fdi_zero_w} are equivalent to
    \begingroup
    \renewcommand*{\arraycolsep}{2pt}
    \begin{equation}\label{eq:harm_proof_fdi}
    \begin{split}
        \sum_{n \in \Z} (\star)^* \mat{\sum^N_{l=1} \tau_l \tilde{\Pi}_l(j \omega n) & 0  \\ 0& -\Pi_\mathrm{QSR}  } \\ \times \mat{P_\mathrm{dd}(j \omega n) & P_\mathrm{dw}(j \omega n) \\ I & 0 \\ P_\mathrm{zd}(j \omega n) & P_\mathrm{zw}(j \omega n) \\ 0 & I} \mat{\hat{\tilde{w}}_{\mathrm{d},n} \\ \hat{w}_n} \leq 0,
    \end{split}
    \end{equation}
    \endgroup
    for all $(\tilde{w}_\mathrm{d}^\top \ w^\top)^\top \in L_2^m[0,T] \times \overline{\mathcal{U}}_\omega^p$, which follows from inserting the relation
    \begingroup
    \begin{equation}\label{eq:zero_w_identity}
        \mat{P_\mathrm{dd}(j \omega n) & P_\mathrm{dw}(j \omega n) \\ I & 0 \\ P_\mathrm{zd}(j \omega n) & P_\mathrm{zw}(j \omega n) \\ 0 & I} \mat{\hat{\tilde{w}}_{\mathrm{d},n} \\ 0} = \mat{P_\mathrm{dd}(j \omega n) \\ I  \\ P_\mathrm{zd}(j \omega n)  \\ 0 } \hat{\tilde{w}}_{\mathrm{d},n} ,
    \end{equation}
    \endgroup
    in~\eqref{eq:iqc_fdi_zero_w}, and the fact that $w \in \overline{\mathcal{U}}_\omega^p$ has no DC term, i.e., $\hat{w}_0 = 0$. Via the Fourier transform~\eqref{eq:iqc_harmonic}$\iff$\eqref{eq:iqc_fourier_harmonic},~\eqref{eq:harm_proof_fdi} for all $(\tilde{w}_\mathrm{d}^\top \ w^\top)^\top \in L_2^m[0,T] \times \overline{\mathcal{U}}_\omega^p$ is equivalent to
    \begingroup
    \renewcommand*{\arraycolsep}{2pt}
    \begin{equation}\label{eq:harm_proof_tdi}
    \begin{split}
        \int_0^T (\star)^\top \mat{\sum^N_{l=1} \tau_l \tilde{\Pi}_l & 0  \\ 0& -\Pi_\mathrm{QSR}} \\ \times \mat{\tilde{P}_\mathrm{dd} & \tilde{P}_\mathrm{dw} \\ I & 0 \\ \tilde{P}_\mathrm{zd} & \tilde{P}_\mathrm{zw} \\ 0 & I} \mat{\tilde{w}_\mathrm{d}(\tau) \\ w(\tau)} \dd \tau \leq 0, \quad T = 2 \pi / \omega,
    \end{split}
    \end{equation}
    \endgroup
    for all $(\tilde{w}_\mathrm{d}^\top \ w^\top)^\top \in L_2^m[0,T] \times \overline{\mathcal{U}}_\omega^p$. By using the relations 
    \begin{equation*}
    \begingroup
    \renewcommand*{\arraycolsep}{2pt}
        \mat{\tilde{z} \\ w} = \mat{\tilde{P}_\mathrm{zd} & \tilde{P}_\mathrm{zw} \\ 0 & I} \mat{\tilde{w}_\mathrm{d} \\ w}, \, \mat{\tilde{z}_\mathrm{d} \\ \tilde{w}_\mathrm{d}} = \mat{\tilde{P}_\mathrm{dd} & \tilde{P}_\mathrm{dw} \\ I & 0} \mat{\tilde{w}_\mathrm{d} \\ w},
    \endgroup
    \end{equation*}
    it follows that~\eqref{eq:harm_proof_tdi} is equivalent to
    \begin{equation}\label{eq:harm_proof_s_lemma_tdi}
        \maketextstyle - \sigma^T_{\Pi_\mathrm{QSR}}(w,\tilde{z}) + \sum^N_{l=1} \tau_l \sigma^T_{\tilde{\Pi}_l}(\tilde{w}_\mathrm{d},\tilde{z}_\mathrm{d})  \leq 0.
    \end{equation}
    Finally,~\eqref{eq:harm_proof_s_lemma_tdi} implies $\sigma^T_{\Pi_\mathrm{QSR}}(w,\tilde{z}) \geq 0$ via the trivial direction of the S-lemma~\cite{jonssonLectureNotesIntegral2001}, since $ \sigma^T_{\tilde{\Pi}_l}(\tilde{w}_\mathrm{d},\tilde{z}_\mathrm{d}) \geq 0$ for all $\tilde{w}_\mathrm{d} = \tilde{\Phi} \tilde{z}_\mathrm{d}$, proving QSR performance of $\tilde{\Phi} \star \tilde{P}$ on $\overline{\mathcal{U}}_\omega^p$.
\end{proof}

Since~\eqref{eq:lfr_closed_loop} is assumed to be stable, the FDIs~\eqref{eq:iqc_fdi_harmonic} and~\eqref{eq:iqc_fdi_zero_w} may be non-strict, and the condition $0 \geq Q$ in $\Pi_\mathrm{QSR}$ may be dropped, as opposed to the conditions for Proposition~\ref{prop:L2_performance_iqc}.

When choosing $Q = -I,S=0$, and $R = \gamma^2 I$, i.e., $\Pi_\mathrm{QSR} = \Pi_\gamma$, QSR performance of $\tilde{\Phi} \star \tilde{P} : w \mapsto \tilde{z}$ for $w \in \overline{\mathcal{U}}_\omega^p$ implies $\norm{\tilde{z}}_T \leq \gamma \norm{w}_T$, providing the upper bound $\overline{\gamma}_\omega(\Phi \star P ) \leq \gamma$ in~\eqref{eq:gamma_harmonic}. In Section~\ref{sec:bode_comp_gkyp} below, we will discuss how~\eqref{eq:iqc_fdi_harmonic} and~\eqref{eq:iqc_fdi_zero_w} can be verified using Lemma~\ref{lemma:gkyp}.

\subsection{Computing the Sinusoidal and Subharmonic Gain}\label{sec:sin_subh_gain}

When a sinusoidal input $w \in \mathcal{U}^p_\omega$ is taken and~\eqref{eq:lfr_closed_loop} is period-preserving, the output of the nonlinearity obeys $w_\mathrm{d} \in L_2^m[0,T]$. As $w$ has only one nonzero Fourier coefficient $\hat{w}_1$, while the Fourier coefficients $\hat{w}_{\mathrm{d},n}$ of $w_\mathrm{d}$ may be nonzero for all $n \in \N$, the FDI in~\eqref{eq:iqc_fdi_harmonic} can be simplified as follows.

\begin{theorem}\label{thm:sinusoidal_performance_iqc}
    Suppose that~\eqref{eq:lfr_closed_loop} is well-posed, stable, and period-preserving. For $\omega>0$, let $\tilde{\Pi}_l$ be bounded self-adjoint operators on $L_2^{k+m}[0,T]$ such that the map $\tilde{\Phi}$ satisfies the IQC $\sigma^T_{\tilde{\Pi}_l}(\tilde{\Phi}(z),z) \geq 0$ for all $z \in L_2^k[0,T]$ and $l=1,\dots,N$, and let $Q\in \mathbb{S}^q, S \in \R^{q \times p}$ and $R \in \mathbb{S}^p$. 
    
    If there exist $\tau_l \geq 0$ for $l=1,\dots,N$ such that the FDI in~\eqref{eq:iqc_fdi_harmonic} holds for $\tilde{\omega} = \omega$, and the FDI in~\eqref{eq:iqc_fdi_zero_w} holds for $\tilde{\omega} =\omega n$, where $1 \neq n \in \N$,
    then $\tilde{\Phi} \star \tilde{P} : w \mapsto \tilde{z}$ has QSR performance for $w \in \mathcal{U}_\omega^p$. 
\end{theorem}
\begin{proof}
    From the same reasoning as in the proof of Theorem~\ref{thm:harmonic_performance_iqc}, for all $w \in \mathcal{U}_\omega^p$, there exist unique $T$-periodic steady-state signals $(\tilde{z}^\top \ \tilde{z}_\mathrm{d}^\top \ \tilde{w}_\mathrm{d}^\top)^\top \in L_2^{q+k+m}[0,T]$. By using the relation~\eqref{eq:zero_w_identity}, the FDI~\eqref{eq:iqc_fdi_harmonic} for $\tilde{\omega} = \omega$ together with the FDI~\eqref{eq:iqc_fdi_zero_w} for $\tilde{\omega} = \omega n$ with $1 \neq n \in \N$ are equivalent to~\eqref{eq:harm_proof_fdi} for all $(\tilde{w}_\mathrm{d}^\top \ w^\top)^\top \in L_2^m[0,T] \times \mathcal{U}_\omega^p $. Following the identical reasoning as in the remainder of the proof of Theorem~\ref{thm:harmonic_performance_iqc}, we conclude that $\tilde{\Phi} \star \tilde{P}$ has QSR performance on $\mathcal{U}_\omega^p$.
\end{proof}

Analogously, performance for subharmonic inputs is verified as follows.

\begin{theorem}\label{thm:subharmonic_performance_iqc}
    Suppose that~\eqref{eq:lfr_closed_loop} is well-posed, stable, and period-preserving. For $\omega>0$ and all $M \in \N_+$, let $\tilde{\Pi}_l$ be bounded self-adjoint operators on $L_2^{k+m}[0,T]$, where $T = M \frac{2 \pi}{\omega}$, such that for all $l=1,\dots,N$, the map $\tilde{\Phi}$ satisfies the IQC $\sigma^T_{\tilde{\Pi}_l}(\tilde{\Phi}(z),z) \geq 0$ for all $z \in L_2^k[0, T]$, and let $Q\in \mathbb{S}^q, S \in \R^{q \times p}$ and $R \in \mathbb{S}^p$. 
    
    If there exist $\tau_l \geq 0$ for $l=1,\dots,N$ such that~\eqref{eq:iqc_fdi_harmonic} holds for $ \tilde{\omega} \in [0,\omega]$ and~\eqref{eq:iqc_fdi_zero_w} holds for $\tilde{\omega} \in (\omega, \infty)$, then $\tilde{\Phi} \star \tilde{P} : w \mapsto \tilde{z}$ has QSR performance for $w \in \underline{\mathcal{U}}_\omega^p$.
\end{theorem}
\begin{proof}
    Fix some $w \in \underline{\mathcal{U}}_\omega^p$, then $w$ has period $T = M \frac{2 \pi}{\omega}$ for some $M \in \N_+$. From the same reasoning as in the proof of Theorem~\ref{thm:harmonic_performance_iqc}, there exist unique $T$-periodic steady-state signals $(\tilde{z}^\top \ \tilde{z}_\mathrm{d}^\top \ \tilde{w}_\mathrm{d}^\top)^\top \in L_2^{q+k+m}[0,T]$. Using the relation~\eqref{eq:zero_w_identity} it follows that the FDI~\eqref{eq:iqc_fdi_harmonic} for $\tilde{\omega} = \frac{\omega n}{M}$ with $n = 1,\dots,M$ together with the FDI~\eqref{eq:iqc_fdi_zero_w} for $\tilde{\omega} = 0$ and $\tilde{\omega} = \frac{\omega(M+n)}{M}$ with $n \in \N_+$ imply~\eqref{eq:harm_proof_fdi} for the fixed $w \in \underline{\mathcal{U}}_\omega^p$, with corresponding steady-state $\tilde{w}_\mathrm{d} \in L_2^m[0,T]$. The identical reasoning as in the remainder of the proof of Theorem~\ref{thm:harmonic_performance_iqc} yields $\sigma^T_{\Pi_\mathrm{QSR}}(w,\tilde{z}) \geq 0$ for the $w \in \underline{\mathcal{U}}_\omega^p$ we fixed, hence $\tilde{\Phi} \star \tilde{P}$ has QSR performance on $\underline{\mathcal{U}}_\omega^p$.
\end{proof}

\subsection{Computation of the Bode Diagram using the GKYP Lemma}\label{sec:bode_comp_gkyp}

The FDIs in Theorems~\ref{thm:harmonic_performance_iqc},~\ref{thm:sinusoidal_performance_iqc} and~\ref{thm:subharmonic_performance_iqc} are all infinite-dimensional. To verify these in a computationally tractable fashion, we will use Lemma~\ref{lemma:gkyp}. For this, we rewrite the FDI in~\eqref{eq:iqc_fdi_harmonic} with $\Pi_\mathrm{QSR}=\Pi_\gamma$ as 
\begin{equation}\label{eq:fdi_g_p}
    (\star)^* \mathcal{P} \tilde{G}(j \tilde{\omega}) \leq 0, \quad \mathcal{P} := \mat{\sum^N_{l=1} \tau_l \tilde{M}_l & 0  \\ 0& -\Pi_\gamma  }, 
\end{equation}
where $\tilde{\Pi}_l = \tilde{\Psi}^* \tilde{M}_l \tilde{\Psi}$ with shared $\tilde{\Psi} \in RH_\infty^{v \times (k+m)}$, and $\tilde{G} \in RH_\infty^{(v+q+p)\times (m+p)}$ contains all the dynamics of $P$ and $\tilde{\Psi}$ combined. Similarly,~\eqref{eq:iqc_fdi_zero_w} is written as $(\star)^* \mathcal{P} \tilde{G}_\mathrm{d}(j \tilde{\omega}) \leq 0$, where $\tilde{G}_\mathrm{d} \in RH_\infty^{(v+q+p)\times m}$. For the remainder of this section, let $(A,B,C,D)$ be a controllable state-space representation of $\tilde{G}$, then $(\star)^* \mathcal{P} \tilde{G}(j \omega) = (\star)^* M_\gamma \mat{(j \omega I - A)^{-1} B \\ I} $ with $M_\gamma := (\star)^* \mathcal{P} (C \ D)$, which is of the form of~\eqref{eq:gkyp_fdi}.

Using the GKYP lemma, we obtain the following semidefinite program for bounding\footnote{The hat distinguishes the \emph{definition} of $\gamma,\overline{\gamma},\underline{\gamma}$ in~\eqref{eqs:gammas} from their \emph{upper bounds} $\hat{\gamma},\hat{\overline{\gamma}},\hat{\underline{\gamma}}$.}~\eqref{eq:gamma_harmonic} using Theorem~\ref{thm:harmonic_performance_iqc}. 
\begin{equation}\label{eq:gamma_bound_harm}
\begin{aligned}
    \hat{\overline{\gamma}}_\omega(\Phi \star P) & := \min_{\gamma^2,\tau_l,P,Q} \quad \gamma, \\
    \text{subject to} \quad & \gamma^2 \geq 0, \quad Q \geq 0, \\
    & \tau_l \geq 0 \text{ for } l=1,\dots,N,  \\
    & \eqref{eq:gkyp_lmi} \text{ with } \varpi = \omega, M = M_\gamma, \\
    & \eqref{eq:iqc_fdi_zero_w} \text{ with } \tilde{\omega} = 0.
\end{aligned}
\end{equation}
Note that since $\{n \omega \mid n \in \N_+\} \subsetneq [\omega, \infty]$, the GKYP lemma is potentially more conservative than~\eqref{eq:iqc_fdi_harmonic}. Similarly, we obtain the following optimization problem to bound~\eqref{eq:gamma_sinus} using Theorem~\ref{thm:sinusoidal_performance_iqc} and the GKYP lemma.
\begin{equation}\label{eq:gamma_bound_sin}
\begin{aligned}
    \hat{\gamma}_\omega(\Phi \star P) & := \min_{\gamma^2,\tau_l,P,Q} \quad \gamma, \\
    \text{subject to} \quad & \gamma^2 \geq 0, \quad Q \geq 0, \\
    & \tau_l \geq 0 \text{ for } l=1,\dots,N,  \\
    & \eqref{eq:iqc_fdi_harmonic} \text{ with } \tilde{\omega} = \omega,\\
    & \eqref{eq:iqc_fdi_zero_w} \text{ with }  \tilde{\omega} = 0,\\
    & \eqref{eq:gkyp_lmi} \text{ with } \varpi = 2\omega, \, M = M_\gamma, \\ \text{and }&A=A_\mathrm{d},B=B_\mathrm{d},C=C_\mathrm{d},D=D_\mathrm{d},
\end{aligned}
\end{equation}
where $(A_\mathrm{d},B_\mathrm{d},C_\mathrm{d},D_\mathrm{d})$ is a controllable state-space representation of $\tilde{G}_\mathrm{d}$.

Finally, bounding~\eqref{eq:gamma_subharmonic} using Theorem~\ref{thm:subharmonic_performance_iqc} requires two separate applications of Lemma~\ref{lemma:gkyp}, each having their own $P,Q$ matrices, but shared $\tau_l$ variables. 
\begin{equation}\label{eq:gamma_bound_sub}
\begin{aligned}
    \hat{\underline{\gamma}}_\omega(\Phi \star P) & := \min_{\gamma^2,\tau_l,P_1,P_2,Q_1,Q_2} \quad \gamma, \\
    \text{subject to} \quad & \gamma^2 \geq 0, \quad Q_1 \geq 0, \quad Q_2 \geq 0,\\
    & \tau_l \geq 0 \text{ for } l=1,\dots,N,  \\
    \eqref{eq:gkyp_lmi_low} \text{ with } & \varpi = \omega, M = M_\gamma, P=P_1,Q=Q_1, \\ 
    \eqref{eq:gkyp_lmi} \text{ with } & \varpi = \omega, M = M_\gamma, P=P_2,Q=Q_2,\\ \text{and }&A=A_\mathrm{d},B=B_\mathrm{d},C=C_\mathrm{d},D=D_\mathrm{d}. 
\end{aligned}
\end{equation}

Even though it is possible to find multipliers $\tilde{\Pi}_l$ that hold for the steady-state map $\tilde{\Phi}$ at specific frequencies $\omega$, in practice one typically selects multipliers $\Pi_l$ for $\Phi : L_2^k \to L_2^m$ which also hold for the steady-state map $\tilde{\Phi} : L_2^k[0,T] \to L_2^m[0,T]$ for all $\omega \in \R_{>0}$ and $T = \frac{2 \pi}{\omega}$, e.g., static multipliers for static maps~\cite{jonssonAnalysisPeriodicallyForced2003}. One first uses these multipliers to verify stability using Proposition~\ref{prop:L2_performance_iqc}. Then, one can use the same state-space representation for the FDI in~\eqref{eq:iqc_fdi} for~\eqref{eq:iqc_fdi_harmonic}, and by removing the input channel $w$, for~\eqref{eq:iqc_fdi_zero_w}.

There are a few immediate extensions of the methods presented in this section. Bounds for the incremental version of~\eqref{eqs:gammas} can be obtained by using IIQCs in~\eqref{eq:gamma_bound_harm},~\eqref{eq:gamma_bound_sin} and~\eqref{eq:gamma_bound_sub}. Additionally, for systems with more than one input channel, one may define even more refined frequency-dependent gains where some channels receive sinusoidal inputs, others harmonic inputs, and the rest subharmonic inputs. The computation of those gains is achieved by a straightforward combination of Theorems~\ref{thm:harmonic_performance_iqc},~\ref{thm:sinusoidal_performance_iqc} and~\ref{thm:subharmonic_performance_iqc}, where the key is to restrict the Fourier content of the input $\hat{w}_n$ as in~\eqref{eq:zero_w_identity} to the adequate input space.

While~\cite{jonssonAnalysisPeriodicallyForced2003} provides useful insights on IQC analysis with $L_2[0,T]$, their approach differs significantly from the one presented in this paper. First, our performance objectives differ substantially from those in~\cite{jonssonAnalysisPeriodicallyForced2003}. We consider $L_2$-gain bounds for the signal spaces in~\eqref{eqs:input_spaces}, while in~\cite{jonssonAnalysisPeriodicallyForced2003} inputs are taken from $L_2^p[0,T]$ and the performance objective is either a supremum norm bound on the output, or gain from a finite set of input harmonics to a finite set of output harmonics. These objectives correspond to performance multipliers that are non-diagonal in the Fourier domain. Second, from a computational viewpoint,~\cite{jonssonAnalysisPeriodicallyForced2003} either uses a frequency theorem for periodic systems in combination with a cutting-plane algorithm, i.e., non-smooth optimization, or relies on the truncation of the Fourier series representation for which the size grows with the number of retained harmonics. In contrast, our computational approach solely relies on semidefinite programming, and gives guaranteed $L_2$-gain bounds since no truncation of the Fourier series is used. Finally, convergence to the periodic solution in~\cite{jonssonAnalysisPeriodicallyForced2003} is established in terms of $L_2[0,T]$-norm convergence. However, no connection is made between the $L_2$-gain including transients, and its steady-state performance on $L_2[0,T]$. We do this via Lemma~\ref{lemma:pp-lemma}, which allows for the interpretation of the gains in~\eqref{eqs:gammas} as a nonlinear Bode diagram.

Finally, we note that the computation of the nonlinear Bode diagrams in Sections~\ref{sec:harmonic_gain} and~\ref{sec:sin_subh_gain} does not require $\tilde{w}_\mathrm{d}$ and $\tilde{z}_\mathrm{d}$ to lie in any of the sets in~\eqref{eqs:input_spaces}. In contrast, these signals were required to have zero DC component in~\cite{krebbekxNonlinearBandwidthBode2025a}, making the method presented in this paper slightly more general.

\section{Application: Control of an Unbalanced Disk}\label{sec:application}

In this section we will demonstrate the use of the nonlinear Bode diagram on a practical controller design analysis task. The setup we consider is an unbalanced disk, and the task is to analyze whether one of two candidate controllers satisfies the given performance requirements. First, we will compute the nonlinear Bode diagram using an ideal model of the system using the methods in Section~\ref{sec:nl_bode_computation}. Here, we demonstrate how the use of ZF multipliers can enable the characterization of integrator behavior. Then, we verify the gain bounds in the Bode diagram in simulation, where we also show what it means if an output performance filter is violated. Finally, we validate the simulation results by repeating them as experiments on the real setup, showing both the theoretical and practical relevance of our approach.

\subsection{System Definition and Control Configuration}

We consider a disk with angular position $\theta$ ($\mathrm{rad}$) and inertia $J=2.4 \cdot 10^{-4}$ ($\mathrm{kg \cdot m^2}$). A mass $M = 7.6 \cdot 10^{-2}$ ($\mathrm{kg}$) is attached to the disk at a distance $l = 4.1 \cdot 10^{-2}$ ($\mathrm{m}$) from the center, resulting in an off-axis center of mass. Given an input $V$ ($\mathrm{V}$), a DC motor applies a torque to the disk, resulting in the differential equation
\begin{equation}\label{eq:ode-disk}
    \frac{\dd^2 \theta}{\dd t^2}(t) = -\frac{M g l}{J} \sin(\theta(t)) - \frac{1}{\tau} \frac{\dd \theta}{\dd t}(t) + \frac{K_\mathrm{m}}{\tau} V(t),
\end{equation}
where $K_\mathrm{m} = 11$ ($\mathrm{rad/Vs}$) is the motor constant, $\tau = 0.40$ ($\mathrm{s}$) is a time constant and $g = 9.81$ ($\mathrm{m/s^2}$) is the gravitational acceleration, and the parameter values are taken from~\cite{koelewijnPhysicalParameterEstimation2019}. The actuator of the disk is limited to voltages $V(t) \in [-10,10]$.

\begin{figure}[t]
    \centering
    \begin{subfigure}[t]{0.73\linewidth}
        \centering
        \includegraphics[width=\linewidth]{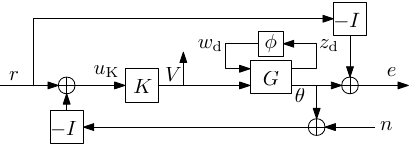}
        \caption{Control configuration of the unbalanced disk.}
        \label{fig:unbalanced_disk_block_diagram}
    \end{subfigure}
    \hfill
    \begin{subfigure}[t]{0.25\linewidth}
        \centering
        \includegraphics[width=\linewidth]{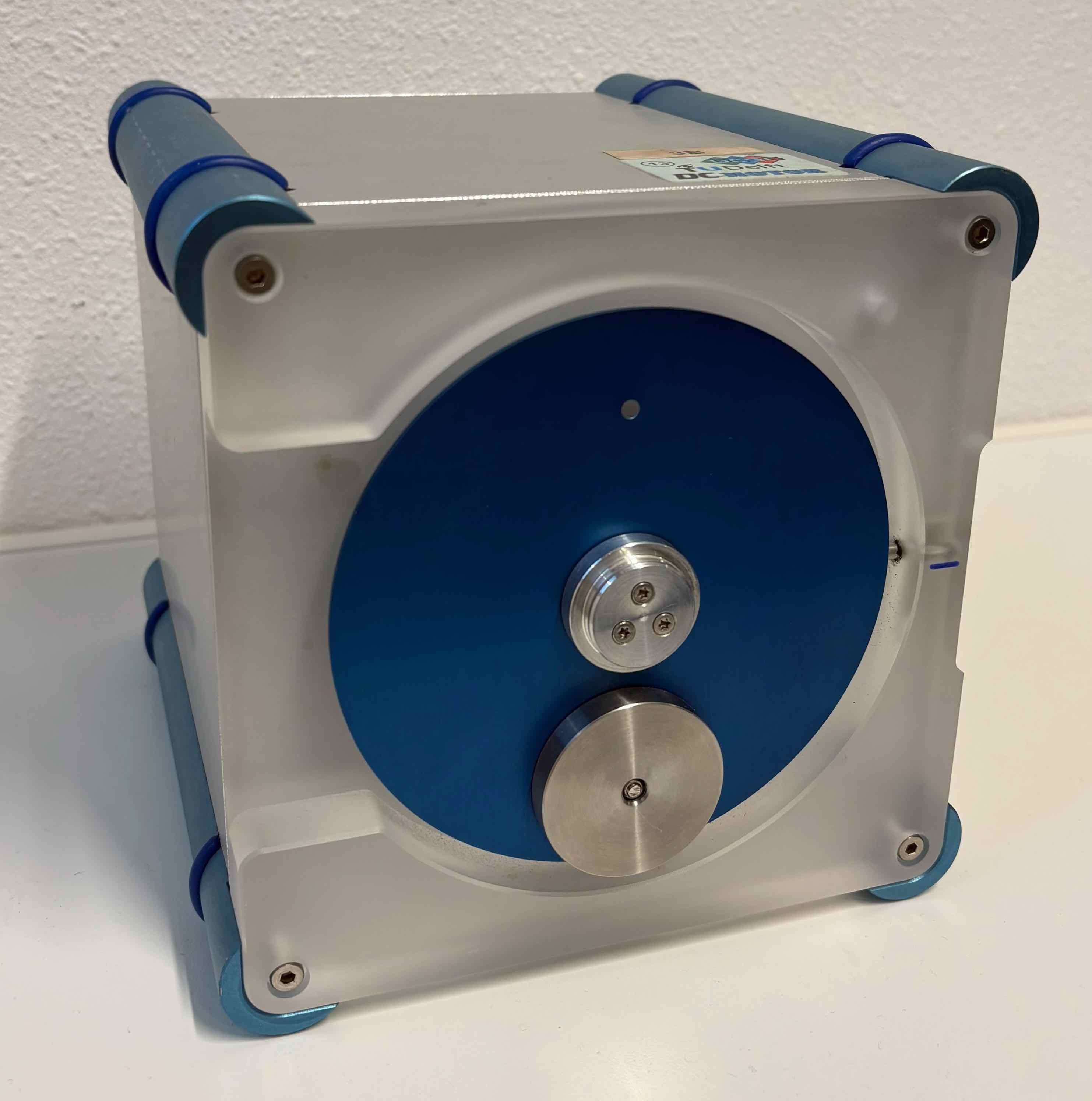}
        \caption{The setup.}
        \label{fig:disk-system}
    \end{subfigure}
    \caption{The unbalanced disk system.}
\end{figure}

We consider the control configuration in Fig.~\ref{fig:unbalanced_disk_block_diagram}, where a reference $r$ is tracked using a feedback controller $K$, and a measurement noise $n$ is taken into account. By defining $w_\mathrm{d} = \phi(z_\mathrm{d})$, $z_\mathrm{d} = \theta$, and $\phi : \theta(t) \mapsto \sin(\theta(t))$,~\eqref{eq:ode-disk} can be written as $\phi \star G$, where $G$ is LTI. The \emph{open-loop} plant $P_\mathrm{ol}$ is the map $P_\mathrm{ol} : (w_\mathrm{d} \ r \ n \ V)^\top \mapsto (z_\mathrm{d} \ e \ V \ u_\mathrm{K})^\top$, which is obtained by disconnecting $\phi$ and the controller $K$ in Fig.~\ref{fig:unbalanced_disk_block_diagram}. If a controller $K : u_\mathrm{K} \mapsto V$ stabilizes $P_\mathrm{ol}$ (see~\cite[Sec. 5.3]{zhouRobustOptimalControl1996} for the definition), then $P := P_\mathrm{ol} \star K \in RH_\infty^{3 \times 3}$ holds, and $\phi \star P : (r \ n)^\top \mapsto (e \ V)^\top$ is the closed loop in Fig.~\ref{fig:unbalanced_disk_block_diagram}.

\subsection{Performance Specification and Controllers}

The expected frequency content of $r,n$ is described with filters $W_\mathrm{r}, W_\mathrm{n}$ such that $r = W_\mathrm{r} \tilde{r}$ and $n = W_\mathrm{n} \tilde{n}$, where $\norm{\tilde{r}}_\mathrm{RMS} \leq 1$ and $\norm{\tilde{n}}_\mathrm{RMS} \leq 1$. The frequency content of $r$ is expected to roll off after $5 \, \mathrm{rad/s}$ with a slope of $-40\mathrm{dB}$ per decade, and the noise $n$ is expected to lie below $-20\mathrm{dB}$ at all frequencies. These specifications are satisfied by\footnote{Using \texttt{makeweight(1,[5,db2mag(-3)],0.01,0,2)} for this second-order continuous-time filter in MATLAB.} $W_\mathrm{r} = \frac{0.01 s^2 + 0.7079 s+25.06}{ s^2 + 7.079 s+25.06}$ and $W_\mathrm{n} = 0.1$, which yield the input filter $W_\mathrm{in} = \mathrm{diag}(W_\mathrm{r} , W_\mathrm{n})$.

The outputs $e,V$ are required to satisfy $\norm{\tilde{e}}_\mathrm{RMS} \leq 1$ and $\norm{\smash{\tilde{V}}}_\mathrm{RMS} \leq 1$, where $\tilde{e} = W_\mathrm{e} e$ and $\tilde{V} = W_\mathrm{V} V$. The tracking error $e$ is required to have a bandwidth of $2 \, \mathrm{rad/s}$, which is satisfied by\footnote{Obtained via \texttt{makeweight(1/.5,[2,1/db2mag(-3)],1)}.} $W_\mathrm{e} = \frac{s+2.818}{s+1.409}$, and we choose $W_\mathrm{V} = \frac{s+40}{s+400}$ to allow the controller to achieve reasonable performance. 

We consider two candidate controllers for these performance requirements, which are filtered PD and PID controllers%
\begin{subequations}\label{eqs:controllers}%
\begin{align}
    \textstyle K_1(s) &= \textstyle \big(7 + \frac{0.5 s}{s/60+1}\big) \frac{1}{s/100+1}, \label{eq:controller_K1} \\
    \textstyle K_2(s) &= \textstyle \big(10 + \frac{20}{s} + \frac{0.8s}{s/100+1}\big)\frac{1}{s/100+1}. \label{eq:controller_K2}
\end{align}
\end{subequations}
In subsequent sections, we will analyze the performance of the controllers in~\eqref{eqs:controllers} using the nonlinear Bode diagram, and verify this performance via simulations and experiments.

When replacing the sinusoidal nonlinearity with a static gain $\Delta \in [-1,1]$, the LTI closed loop $P_{\Delta,i} : =(\Delta \star P_\mathrm{ol} \star K_i)_{e \leftarrow r}$ for the map $r \mapsto e$ has the following integrator behavior. For $i=1$, i.e., $K=K_1$, the closed loop $P_{\Delta,1}$ has a nonzero DC gain for $\Delta \neq 0$, and has bounded $\lim_{s \to 0} \frac{1}{s} P_{\Delta,1}(s)$ only at $\Delta = 0$. On the other hand, $K_2$ has bounded limit $\lim_{s \to 0} \frac{1}{s} P_{\Delta,2}(s)$ for all $\Delta \in [-1,1]$, which guarantees at least one integrator. Since $\theta \mapsto \sin(\theta)$ is static, we may expect an integrator behavior in the low-frequency range in the Bode plot of $r \mapsto e$, i.e., $\lim_{\omega \downarrow 0} \underline{\gamma}_\omega((\phi \star P_\mathrm{ol} \star K_2)_{e \leftarrow r}) = 0$.

\subsection{The Period-Preserving Property}

We must prove that the unbalanced disk system is period-preserving. Otherwise, the nonlinear Bode diagram may not be defined. We achieve this by showing that, for both controllers in~\eqref{eqs:controllers}, the closed loop is exponentially convergent as in~\cite[Lemma 2.37]{pavlovUniformOutputRegulation2006} and~\cite[Sec. III]{pavlovFrequencyDomainPerformance2007}. 

The loop transformed nonlinearity $\chi(x) := x+\phi(x)$ satisfies $\chi'(x) \in [0,2]$ for $x \in \R$, and $\chi \star \mathcal{M} \star P = \phi \star P$, where $\mathcal{M} := \scalebox{.7}{$\mat{0&1 \\ 1&-1}$}$. For each controller, the transformed closed loop $\mathcal{M} \star P$ with the slope condition $\chi'(x) \in [0,2]$ satisfies the conditions in~\cite[Sec. III]{pavlovFrequencyDomainPerformance2007} for exponential convergence for bounded piecewise continuous inputs.

\begin{figure*}[t]
    \centering
    \begin{minipage}[t]{0.49\textwidth}
    \centering
    \begin{subfigure}[t]{\linewidth}
        \centering
        \includegraphics[width=\linewidth]{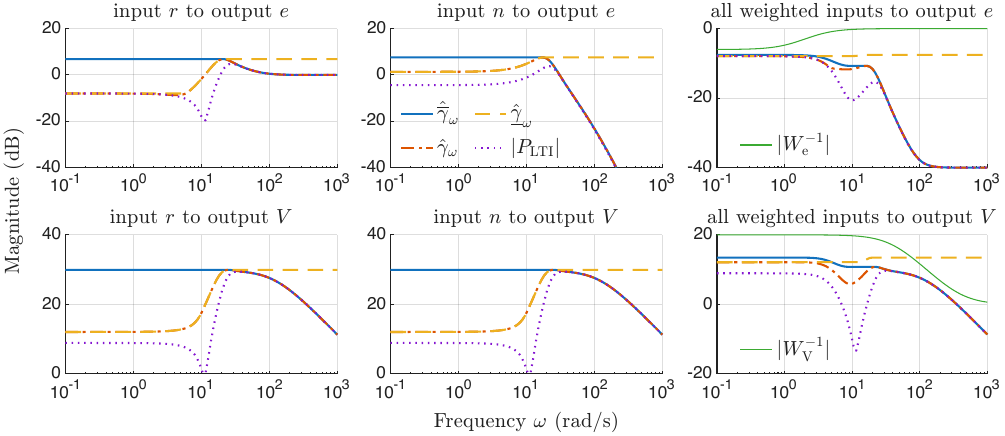}
        \caption{Controller $K_1$ with $\gamma(W_\mathrm{out}(\phi \star P) W_\mathrm{in}) \leq 0.85$.}
        \label{fig:bode_K1_static}
    \end{subfigure}
    \hfill
    \begin{subfigure}[t]{\linewidth}
        \centering
        \includegraphics[width=\linewidth]{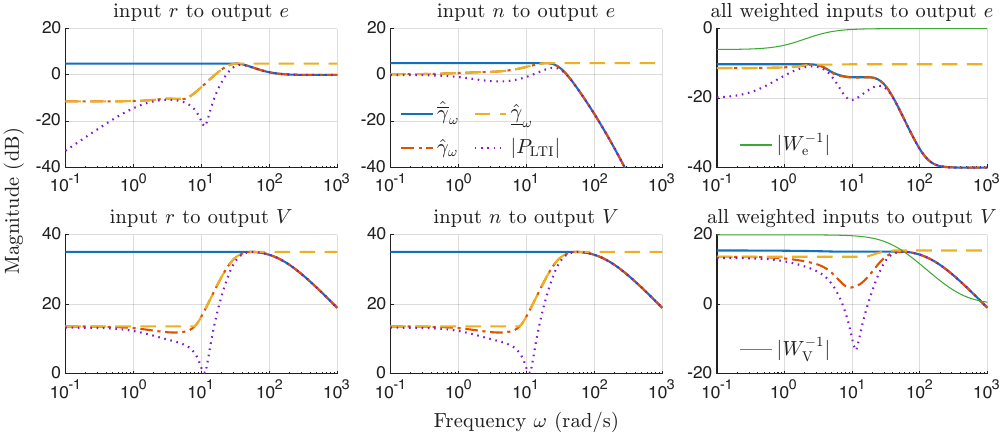}
        \caption{Controller $K_2$ with $\gamma(W_\mathrm{out}(\phi \star P) W_\mathrm{in}) \leq 1.70$.}
        \label{fig:bode_K2_static}
    \end{subfigure}
    \caption{Nonlinear Bode plot using one static multiplier.}
    \label{fig:bodes_disk_static}
    \end{minipage}
    \hfill
    \begin{minipage}[t]{0.49\textwidth}
    \centering
    \begin{subfigure}[t]{\linewidth}
        \centering
        \includegraphics[width=\linewidth]{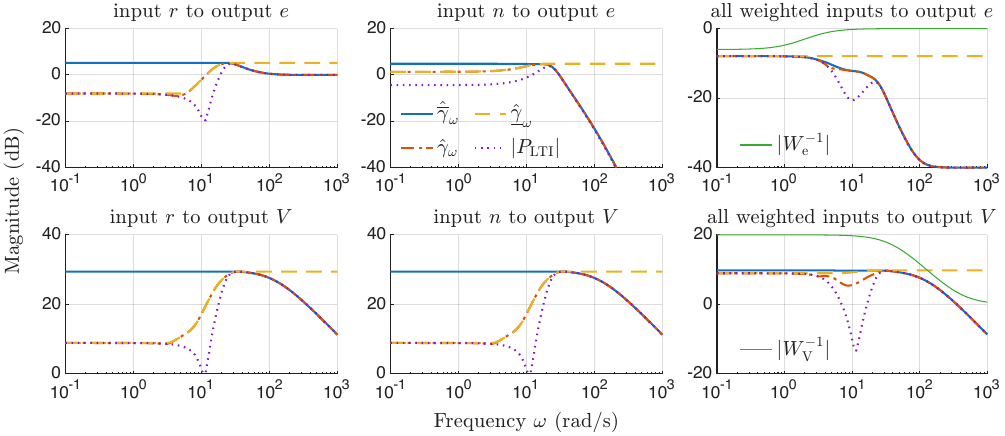}
        \caption{Controller $K_1$ with $\gamma(W_\mathrm{out}(\phi \star P) W_\mathrm{in}) \leq 0.85$.}
        \label{fig:bode_K1_zf}
    \end{subfigure}
    \hfill
    \begin{subfigure}[t]{\linewidth}
        \centering
        \includegraphics[width=\linewidth]{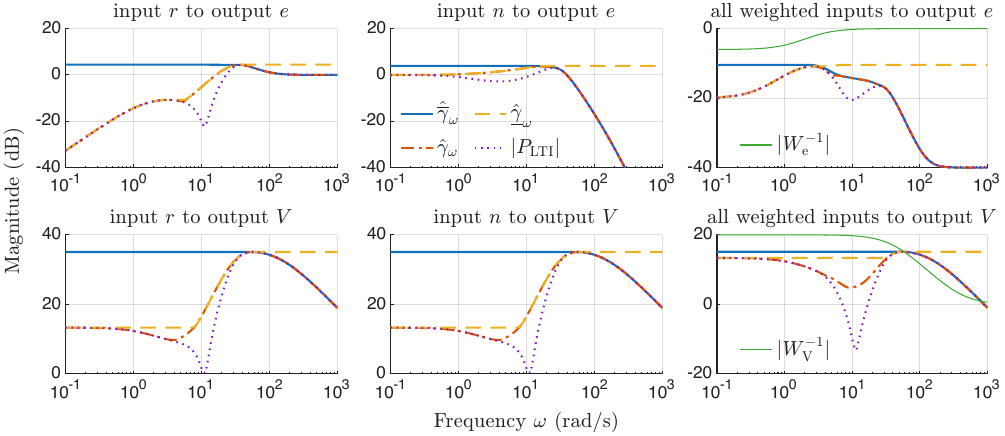}
        \caption{Controller $K_2$ with $\gamma(W_\mathrm{out}(\phi \star P) W_\mathrm{in}) \leq 1.70$.}
        \label{fig:bode_K2_zf}
    \end{subfigure}
    \caption{Nonlinear Bode plot using static and ZF multipliers.}
    \label{fig:bodes_disk_zf}
    \end{minipage}
\end{figure*}

\subsection{Performance Analysis using Nonlinear Bode Diagrams}\label{sec:disk_static_bode}

In order to compute the nonlinear Bode diagram using the upper bounds derived in Section~\ref{sec:nl_bode_computation}, we require IQC multipliers for $\phi$. The nonlinearity satisfies the sector condition $\alpha \leq \sin(\theta)/\theta \leq 1$ for all $\theta \neq 0$, where $\alpha := \min_{\theta \in \R} \sin(\theta)/\theta \approx -0.217$, which is described by $(\phi(z_\mathrm{d}(t)) - \alpha z_\mathrm{d}(t))(z_\mathrm{d}(t) - \phi(z_\mathrm{d}(t))) \geq 0$ for all $t \in \R_{\geq 0}$, and is equivalent to~\cite{schererDissipativityIntegralQuadratic2022}
\begin{equation}\label{eq:sin_mul_static}
    (\star)^\top \Pi_\phi \mat{z_\mathrm{d}(t) \\ \phi(z_\mathrm{d}(t))} \geq 0, \quad \Pi_\phi = \mat{-2 \alpha & 1+\alpha \\ 1+\alpha & -2}.
\end{equation}
Hence, $\phi$ satisfies the IQC $\sigma_{\Pi_\phi}(\phi(z_\mathrm{d}),z_\mathrm{d})  \geq 0$ for all $z_\mathrm{d} \in L_2$ in~\eqref{eq:iqc_def}. Moreover, as both $\phi$ and $\Pi_\phi$ are static, this IQC holds for the steady-state map $\tilde{\phi} : L_2[0,\frac{2 \pi}{\omega}] \to L_2[0,\frac{2 \pi}{\omega}]$ as well, for every $\omega > 0$, such that $\Pi_1=\Pi_\phi$ with $N=1$ can be used in Proposition~\ref{prop:L2_performance_iqc}, and $\tilde{\Pi}_1 = \Pi_1$ with $N=1$ in Theorems~\ref{thm:harmonic_performance_iqc}-\ref{thm:subharmonic_performance_iqc}.

For each controller in~\eqref{eqs:controllers}, the Bode diagram in Fig.~\ref{fig:bodes_disk_static} is computed as follows. For the SISO maps from $r,n$ to $e,V$ (first two columns), and for the maps from both filtered inputs $(\tilde{r} \ \tilde{n})^\top$ to $e$ and to $V$, i.e., $(\phi \star P)_j W_\mathrm{in}$ (third column), we compute the harmonic, sinusoidal and subharmonic gain bounds $\hat{\overline{\gamma}}_\omega$, $\hat{\gamma}_\omega$ and $\hat{\underline{\gamma}}_\omega$ by solving~\eqref{eq:gamma_bound_harm},~\eqref{eq:gamma_bound_sin} and~\eqref{eq:gamma_bound_sub}, respectively, on a grid of $100$ logarithmically spaced frequencies $\omega \in [10^{-1}, 10^3]$ rad/s\footnote{For numerical conditioning, the state-space realizations are time-scaled by a factor $10$ prior to solving the LMIs, which is undone in the plots.}. For comparison, Fig.~\ref{fig:bodes_disk_static} also shows the maximum singular value of the linearization $P_\mathrm{LTI}$, corresponding to $\phi=I$, and the inverse output weights $|W_\mathrm{e}^{-1}|$ and $|W_\mathrm{V}^{-1}|$, which are the frequency-wise bounds required by the performance specification. Finally, an $L_2$-gain bound for the weighted closed loop $W_\mathrm{out}(\phi \star P) W_\mathrm{in}$, with $W_\mathrm{out} = \mathrm{diag}(W_\mathrm{e},W_\mathrm{V})$, is obtained via Proposition~\ref{prop:L2_performance_iqc}.

When comparing $K_1$ and $K_2$ using Fig.~\ref{fig:bodes_disk_static}, one can see that both controllers achieve the desired RMS gain for the error channel, where $K_2$ achieves the best tracking performance. Since $W_\mathrm{e}^{-1}$ is monotonically increasing as function of frequency, one must compare it against the sinusoidal or subharmonic gain. First, the top left plots of Figs.~\ref{fig:bode_K1_static} and~\ref{fig:bode_K2_static} show that $\hat{\underline{\gamma}}_\omega((\phi \star P)_{e \leftarrow r}) \leq -3\mathrm{dB}$ for $\omega \leq 2 \, \mathrm{rad/s}$, which satisfies the performance requirement. Second, the top right plot shows the same property for inputs with expected frequency content since $\hat{\underline{\gamma}}_\omega((\phi \star P)_e W_\mathrm{in}) \leq |W_\mathrm{e}^{-1} (j \omega)|$ for all $\omega \in [10^{-1}, 10^3]$. The expected integrator behavior for $K_2$, however, is not visible in Fig.~\ref{fig:bode_K2_static}.

Then, since $W_\mathrm{V}^{-1}$ is monotonically decreasing, one must compare it with the harmonic gain. The bottom right plot in Figs.~\ref{fig:bode_K1_static} and~\ref{fig:bode_K2_static} shows that $\hat{\overline{\gamma}}_\omega((\phi \star P)_V W_\mathrm{in}) \leq |W_\mathrm{V}^{-1} (j \omega)|$ for all $\omega \in \R_{\geq 0}$ is only satisfied for $K_1$. This agrees with the observation that the weighted closed-loop performance with $K_1$ is below one, while for $K_2$ the bound is larger than one. 

Finally, we note that when using an IIQC for $\phi$, where $-1 \leq \frac{\sin(x_1) - \sin(x_2)}{x_1-x_2} \leq 1$ for all $x_1,x_2 \in \R$ yields $\alpha=-1$ in $\Pi_\phi$, the incremental $L_2$-gain performance of $K_1$ becomes $\Gamma(W_\mathrm{out}(\phi \star P) W_\mathrm{in}) \leq 4.19$, while for $K_2$ one obtains $\Gamma(W_\mathrm{out}(\phi \star P) W_\mathrm{in}) \leq 1.70$. This means that, in the incremental setting, $K_2$ may perform better than $K_1$.

\subsection{Integrator Behavior via Zames-Falb Multipliers}\label{sec:zf_multipliers}

The multiplier $\Pi_\phi$ in~\eqref{eq:sin_mul_static} only uses the sector of $\phi$, which also includes time-varying gains in that sector, and other dynamic nonlinearities. Since a time-varying gain $\phi$ can destroy the integrator behavior of the closed loop with controller $K_2$, it is not surprising that integrator behavior is not present in Fig.~\ref{fig:bode_K2_static} for the map $r \mapsto e$.

Conservatism can be reduced by also exploiting that $\phi$ is an odd static nonlinear map with slope $\phi'(\theta) = \cos(\theta) \in [-1,1]$, using ZF multipliers~\cite{zamesStabilityConditionsSystems1968}. Since ZF multipliers are only valid for \emph{static} functions, they rule out time-varying gains or nonlinearities. Since ZF multipliers are typically defined for odd monotone nonlinearities with no specific slope conditions, see~\cite{zamesStabilityConditionsSystems1968} and~\cite[Sec. 4]{turnerZamesFalbMultipliersDont2022}, we define the transformation
\begin{equation}\label{eq:zf_signals}
    u := T_u \mat{z_\mathrm{d} \\ w_\mathrm{d}}, \quad a := T_a \mat{z_\mathrm{d} \\ w_\mathrm{d}},
\end{equation}
where $T_u := \frac{1}{2} \mat{1 & 1}$ and $T_a := \frac{1}{2} \mat{1 & -1}$. The transformation in~\eqref{eq:zf_signals} is chosen such that the map $\tilde{\psi} : a \mapsto u$ is monotone and odd. This can be seen by noting that $u = \psi(z_\mathrm{d})$, where $\psi(\theta) := (\theta + \sin(\theta))/2$, is odd and monotone with slope $\psi' \in [0,1]$, and using a loop transformation, see~\cite[Sec. 4.1]{turnerZamesFalbMultipliersDont2022}. 

Therefore, $w_\mathrm{d} = \phi(z_\mathrm{d})$ with $\phi'(z_\mathrm{d}) \in [-1,1]$ and~\eqref{eq:zf_signals} obeys $\inner{a}{(m_0 - H)u} \geq 0$ for all $z_\mathrm{d} \in L_2$ and any, possibly noncausal, LTI $H$ with impulse response $h$ satisfying $\norm{h}_1 \leq m_0$~\cite{zamesStabilityConditionsSystems1968}. We parametrize $H$ using the basis $H_i(s) = (p_i/(s+p_i))^{n_i}$, $i=1,\dots,n_H$, whose impulse responses satisfy $\norm{h_i}_1 = H_i(0) = 1$. Hence, for $\sigma \in \{-1,1\}$,
\begin{equation}\label{eq:zf_iqcs}
    \inner{a}{(1-\sigma H_i)u} \geq 0, \quad \inner{(1-\sigma H_i)a}{u} \geq 0, 
\end{equation}
where the second IQC is the anticausal ZF multiplier $1 - \sigma H_i^*$, realized causally by filtering $a$ instead of $u$. A conic combination of~\eqref{eq:zf_iqcs} yields $\inner{a}{Mu} \geq 0$ with $M = m_0 - \sum_{i} (c_i H_i + d_i H_i^*)$ and $\sum_i (|c_i| + |d_i|) \leq m_0$, i.e., the search over $\tau_l$ in~\eqref{eq:gamma_bound_harm}--\eqref{eq:gamma_bound_sub} is a convex search over all ZF multipliers with dynamics in the span of $\{h_i(t),h_i(-t)\}_{i=1}^{n_H}$.

All dynamics are collected in one filter $\Psi$ with output $v = (z_\mathrm{d}, w_\mathrm{d}, a, u, H_1 u, H_1 a, \dots, H_{n_H} u, H_{n_H} a)^\top$. With $E_{xy}$ the symmetric selector matrix such that $v^\top E_{x,y} v = 2xy$, the sector IQC is included as $M_1 = \mathrm{diag}(\Pi_\phi,0)$, and~\eqref{eq:zf_iqcs} as $M_l = E_{a,u} - \sigma E_{a,H_i u}$ and $M_l = E_{a,u} - \sigma E_{H_i a, u}$, giving $N = 1 + 4 n_H$ multipliers. We use $p_i \in \{10^{-2},10^{-1},\dots,10^{2}\}$ rad/s and $n_i \in \{1,2\}$, so $n_H = 10$ and $\Psi$ has $30$ states.

The homotopy condition of Proposition~\ref{prop:L2_performance_iqc} holds, since $(\theta + \tau \sin(\theta))/2$ is odd with slope in $[0,1]$ for all $\tau \in [0,1]$. Moreover, the ZF IQC also holds on the space of periodic signals~\cite[Ex.~3]{jonssonAnalysisPeriodicallyForced2003}. Hence,~\eqref{eq:zf_iqcs} holds on $L_2[0,\frac{2 \pi}{\omega}]$, and can be used in Theorems~\ref{thm:harmonic_performance_iqc}-\ref{thm:subharmonic_performance_iqc}.

The resulting nonlinear Bode diagrams are given in Fig.~\ref{fig:bodes_disk_zf}. First, we note that the $L_2$-gain bound for the weighted closed loop is the same as the bound obtained via static multipliers only, for both controllers. However, frequency-wise, the ZF multipliers show a significantly less conservative bound. This is most notably seen in the $r \mapsto e$ tile for the controller $K_2$, where the integrator behavior now closely matches the linearization, which is the main advantage of using ZF multipliers for the analysis of this system. Other small differences are lower maximum values of the gain in all unweighted diagrams, for both $K_1$ and $K_2$, as well as some small differences between $1$ and $10 \, \mathrm{rad/s}$ in the tiles $r \mapsto V$ and $n \mapsto V$ for $K_2$.

\subsection{Simulation Results}\label{sec:disk_sims}

\begin{figure*}[!t]
    \centering
    \begin{minipage}[t]{0.49\textwidth}
        \centering
        \begin{subfigure}[t]{\linewidth}
            \centering
            \includegraphics[width=\linewidth]{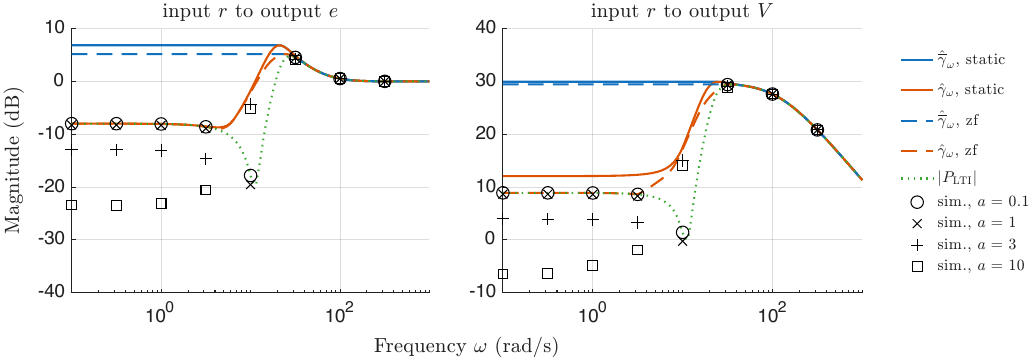}
            \caption{Controller $K_1$.}
            \label{sins_compare_sim_K1}
        \end{subfigure}
        \hfill
        \begin{subfigure}[t]{\linewidth}
            \centering
            \includegraphics[width=\linewidth]{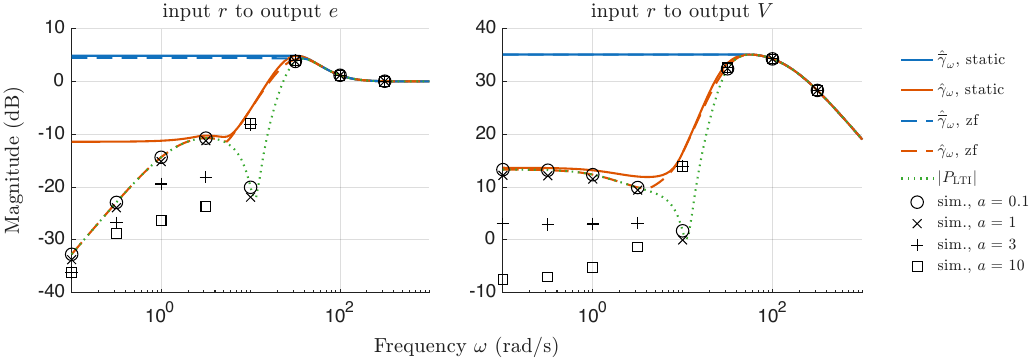}
            \caption{Controller $K_2$.}
            \label{fig:sins_compare_sim_K2}
        \end{subfigure}
        \caption{RMS gain of simulated sinusoidal inputs for different amplitudes vs. the harmonic and sinusoidal gain.}
        \label{fig:sins_compare_sim}
    \end{minipage}
    \hfill
    \begin{minipage}[t]{0.49\textwidth}
        \centering
        \begin{subfigure}[t]{\linewidth}
            \centering
            \includegraphics[width=\linewidth]{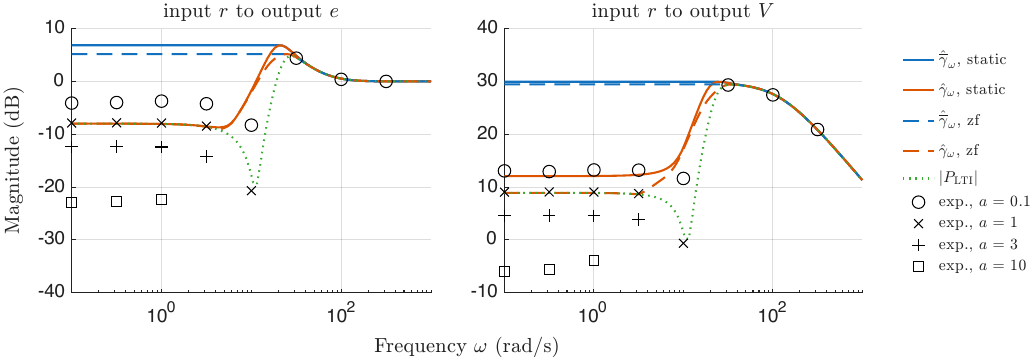}
            \caption{Controller $K_1$.}
            \label{sins_compare_exp_K1}
        \end{subfigure}
        \hfill
        \begin{subfigure}[t]{\linewidth}
            \centering
            \includegraphics[width=\linewidth]{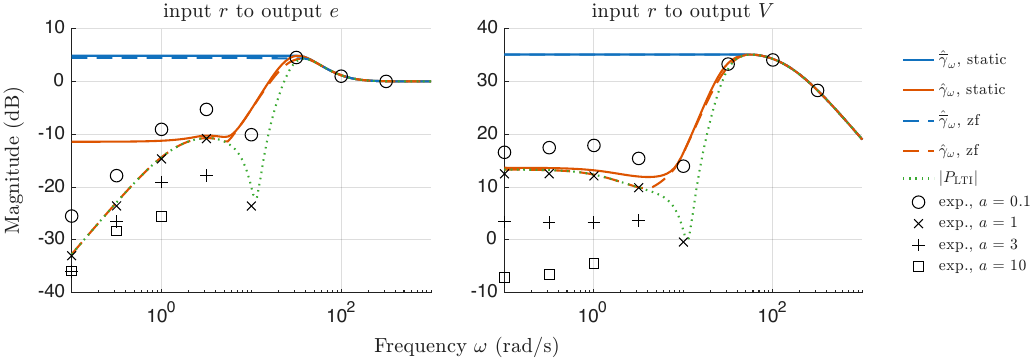}
            \caption{Controller $K_2$.}
            \label{fig:sins_compare_exp_K2}
        \end{subfigure}
        \caption{RMS gain of experimental data from sinusoidal inputs for different amplitudes vs. the harmonic and sinusoidal gain.}
        \label{fig:sins_compare_exp}
    \end{minipage}
\end{figure*}

\begin{figure}[!t]
    \centering
    \begin{subfigure}[t]{.24\linewidth}
        \centering
        \includegraphics[width=\linewidth]{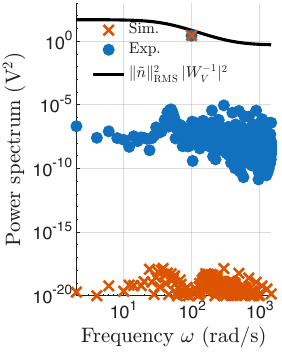}
        \caption{$K_1$, $a=0.1$.}
        \label{fig:ps_compare_K1_a01}
    \end{subfigure}
    \hfill
    \begin{subfigure}[t]{.24\linewidth}
        \centering
        \includegraphics[width=\linewidth]{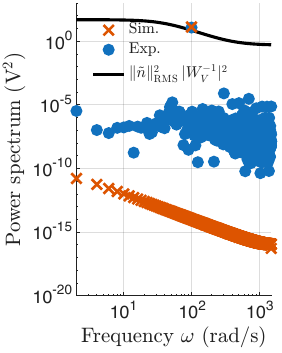}
        \caption{$K_2$, $a=0.1$.}
        \label{fig:ps_compare_K2_a01}
    \end{subfigure}
    \hfill
    \begin{subfigure}[t]{.24\linewidth}
        \centering
        \includegraphics[width=\linewidth]{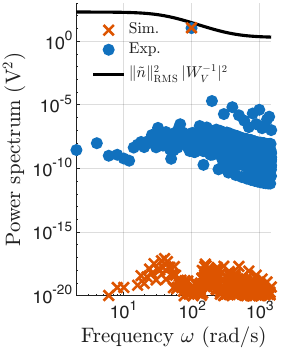}
        \caption{$K_1$, $a=0.2$.}
        \label{fig:ps_compare_K1_a02}
    \end{subfigure}
    \hfill
    \begin{subfigure}[t]{.24\linewidth}
        \centering
        \includegraphics[width=\linewidth]{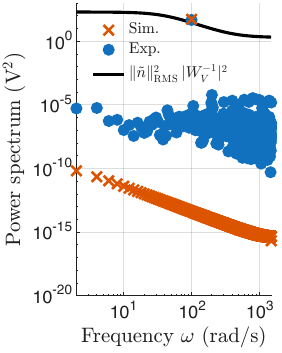}
        \caption{$K_2$, $a=0.2$.}
        \label{fig:ps_compare_K2_a02}
    \end{subfigure}
    \caption{Power spectra of $n \mapsto V$ with $n(t) = a \sin(100t)$, both in simulation and experiment, and the controllers in~\eqref{eqs:controllers}.}
    \label{fig:ps_compare}
\end{figure}

\begin{figure}[t]
    \centering
    \begin{subfigure}[t]{.48\linewidth}
        \centering
        \includegraphics[width=\linewidth]{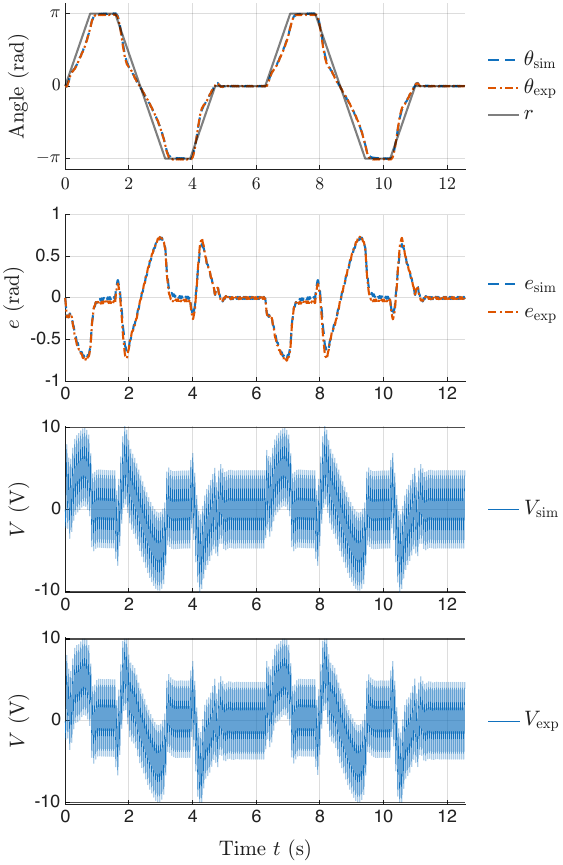}
        \caption{Controller $K_1$.}
        \label{fig:time_sim_exp_K1}
    \end{subfigure}
    \hfill
    \begin{subfigure}[t]{.48\linewidth}
        \centering
        \includegraphics[width=\linewidth]{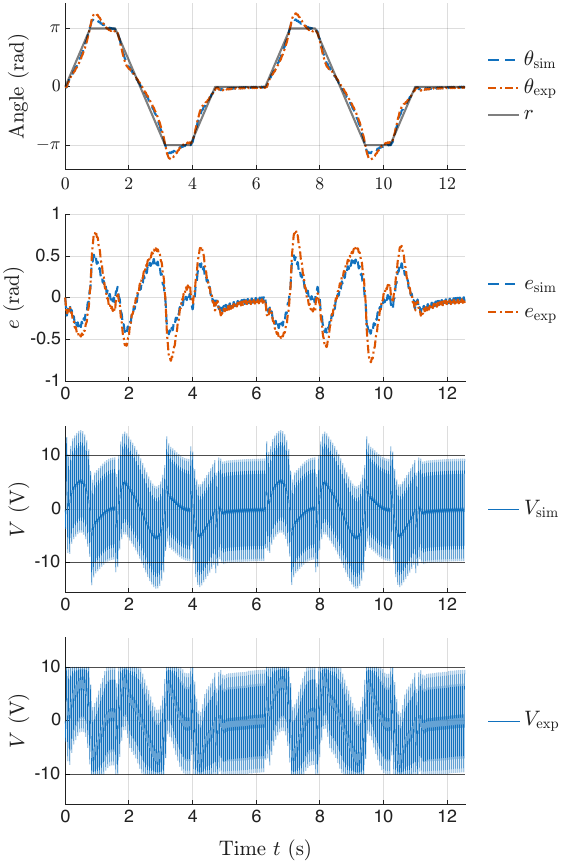}
        \caption{Controller $K_2$.}
        \label{fig:time_sim_exp_K2}
    \end{subfigure}
    \caption{Simulation and experiment for tracking a trapezoidal setpoint with noise $n(t) = 0.2 \sin(100t)$.}
    \label{fig:time_sim_exp}
\end{figure}

To validate the use of our nonlinear Bode diagram for performance analysis, we will perform three different simulations that each cover a different aspect of performance. 

First, we simulate the unbalanced disk model~\eqref{eq:ode-disk} for sinusoidal reference signals $r(t) = a \sin(\omega t)$ with zero noise $n$. We consider the controllers in~\eqref{eqs:controllers}, amplitudes $a \in \{0.1,1,3,10\}$, and 8 logarithmically spaced frequencies $\omega$ in $[10^{-1}, 10^{2.5}]$. For these inputs, the RMS norm is $\norm{r}_\mathrm{RMS} = a/\sqrt{2}$. The RMS norm of the output is computed\footnote{The ordinary differential equation in~\eqref{eq:ode-disk} is solved using MATLAB's \texttt{ode15s} function. Then, the solution is resampled on a uniform grid using MATLAB's \texttt{deval} function. Therefore, the simulation contains no discretization effects (e.g., ZOH) of the controller and/or input $(r \ n)^\top$.} over 10 periods, after the transients have decayed\footnote{Convergence is attained after period $k$ if the RMS norm of period $k+1$ differs less than $0.1\%$ with that of period $k$, where at least $k \geq 20$ is used.}, with 200 samples per period. The resulting RMS gains for $r \mapsto e$ and $r \mapsto V$ are plotted in Fig.~\ref{fig:sins_compare_sim} for both controllers, together with the corresponding sinusoidal and harmonic gains, computed using both static and ZF multipliers. From Fig.~\ref{fig:sins_compare_sim}, it is clear that the sinusoidal gain $\gamma_\omega(\phi \star P)$ is approximated very closely at all tested frequency points by at least one of the sinusoidal inputs, for both controllers. Importantly, all RMS gains from simulation data are upper bounded by the harmonic and sinusoidal gain bounds computed in Sections~\ref{sec:disk_static_bode} and~\ref{sec:zf_multipliers}. Additionally, Fig.~\ref{fig:sins_compare_sim_K2} verifies that the integrator behavior for $K_2$ is correctly quantified by the ZF multipliers in Section~\ref{sec:zf_multipliers}.

Second, we show that for $K_2$, the performance of $n \mapsto V$ violates the specification set by $W_\mathrm{V}$, while for $K_1$ the specification is met. We again simulate~\eqref{eq:ode-disk} for both controllers in~\eqref{eqs:controllers}, now with reference $r=0$ and noise $n(t) = a \sin(100t)$, where $a \in \{0.1, 0.2\}$. We then compute the Fourier coefficients $\hat{V}_n$ of the output $V$, see~\eqref{eq:rms_fourier_L2} using the discrete Fourier transform on 50 periods with 30 samples per period, resulting in a sampling time of $\frac{2 \pi}{3000} \approx 2.1 \, \mathrm{ms}$. The \emph{Power Spectrum} (PS) is computed from the Fourier coefficients as $\mathrm{PS}_V(n) = |\hat{V}_n|^2 + |\hat{V}_{-n}|^2$ for $n \in \N_+$ and $\mathrm{PS}_V(0) = |\hat{V}_0|^2$, where $\mathrm{PS}_V(n)$ is the power of $V$ in mode $n$, and $\norm{V}^2_\mathrm{RMS} = \sum_{n \in \N} \mathrm{PS}_V(n)$ is the total power in $V$. To check if $n \mapsto V$ satisfies the performance specified by $W_\mathrm{V}$, we plot the PS of $V$ together with $\norm{n}^2_\mathrm{RMS} |W_\mathrm{V}^{-1}(j \omega)|^{2}$. If $\mathrm{PS}_V(n)$ exceeds $\norm{n}^2_\mathrm{RMS} |W_\mathrm{V}^{-1}(j \omega_0 n)|^{2}$, where $\omega_0 = 2$ is the fundamental frequency of the Fourier transform, then the specification is violated at that output frequency. The resulting power spectra are plotted in Fig.~\ref{fig:ps_compare}. Note that $\mathrm{PS}_V(n)$ for $K_1$ lies below the performance specified by $W_\mathrm{V}$ in Figs.~\ref{fig:ps_compare_K1_a01} and~\ref{fig:ps_compare_K1_a02}, while the filter specification is violated by $K_2$ at $100 \, \mathrm{rad/s}$ in Figs.~\ref{fig:ps_compare_K2_a01} and~\ref{fig:ps_compare_K2_a02}. This agrees with the performance description in the Bode plots in Figs.~\ref{fig:bodes_disk_static} and~\ref{fig:bodes_disk_zf}; the harmonic gain for $K_1$ in the bottom right plot of Fig.~\ref{fig:bode_K1_static} (or Fig.~\ref{fig:bode_K1_zf}) does not exceed $|W_\mathrm{V}|^{-1}$, while in Fig.~\ref{fig:bode_K2_static} (or Fig.~\ref{fig:bode_K2_zf}), the harmonic gain for $K_2$ does \emph{not} lie below the filter specification. Moreover, since this violation of $|W_\mathrm{V}|^{-1}$ occurs at $100 \, \mathrm{rad/s}$, it is expected that a noise with this frequency content violates the filter specification, which is verified by the power spectra in Fig.~\ref{fig:ps_compare}.

Third, we show the practical difference between the weighted performance $\gamma(W_\mathrm{out}(\phi \star P) W_\mathrm{in}) \leq 0.85$ for $K_1$, and $\gamma(W_\mathrm{out}(\phi \star P) W_\mathrm{in}) \leq 1.70$ for $K_2$, which violates the requirements for $n \mapsto V$. We consider a periodic trapezoidal reference  signal $r$ that starts at zero, ramps linearly between $\pi$ and $-\pi$, with some rests at $0$ in between repetitions. This reference is displayed in gray in the top plot of Fig.~\ref{fig:time_sim_exp}. Additionally, we consider the noise signal $n(t) = 0.2 \sin(100 t)$. Note that $\norm{\tilde{r}}_\mathrm{RMS} = \norm{W_\mathrm{r}^{-1} r}_\mathrm{RMS} \approx 2.13$ and $\norm{\tilde{n}}_\mathrm{RMS} = \norm{W_\mathrm{n}^{-1} n}_\mathrm{RMS} = \sqrt{2} \approx 1.41$. When looking at the third row in Fig.~\ref{fig:time_sim_exp}, it is clear that $K_1$ respects the $\pm 10 \mathrm{V}$ range of the disk's actuator, while $K_2$ yields inputs that exceed the maximum value of the actuator. The latter corresponds to the violation of the filter $|W_\mathrm{V}|^{-1}$ in the bottom right plot of Figs.~\ref{fig:bode_K2_static} and~\ref{fig:bode_K2_zf}.

\subsection{Experimental Results}\label{sec:disk_exps}

In this section, we repeat the analysis from Section~\ref{sec:disk_sims} as experiments on the disk setup in Fig.~\ref{fig:disk-system}. The model in~\eqref{eq:ode-disk} does not capture every aspect of the real setup, such as saturation of the input voltage at $\pm 10 \mathrm{V}$, static friction, and controller discretization. Therefore, some differences between simulation and experiment are expected. 

First, we use a ZOH discretization of the controller and a sampling time between $1$-$5\mathrm{ms}$ such that an integer number of samples fit in one period of $r$. The RMS gains for the sinusoidal inputs are computed from the experimental data in the same way as in Section~\ref{sec:disk_sims}, and they are plotted in Fig.~\ref{fig:sins_compare_exp}, together with the same harmonic and sinusoidal gain bounds as in Fig.~\ref{fig:sins_compare_sim}. For those inputs that yielded voltages exceeding $\pm 10 \mathrm{V}$, the RMS gain was not computed, and hence the marker for that combination of frequency and amplitude is absent in Fig.~\ref{fig:sins_compare_exp}. Except for the amplitude $a=0.1$ for the lowest $4$ frequencies, the RMS gains from the experiment are upper bounded by the sinusoidal gain bounds computed in Sections~\ref{sec:disk_static_bode} and~\ref{sec:zf_multipliers}, which verifies the correctness of our nonlinear Bode diagram in practice. At the low amplitude $a=0.1$, it was apparent that the static friction prevented the disk from moving most of the time, which results in a degraded tracking performance. This could explain the observation that the RMS gain for the first $4$ frequencies at $a=0.1$ exceed the sinusoidal gain bound.

Second, the results of the experiments with $r(t)=0$ and $n(t) = a \sin(100t)$, where $a \in \{0.1,0.2\}$, are plotted together with the simulation results in Fig.~\ref{fig:ps_compare}. We note that the $100 \, \mathrm{rad/s}$ component matches the simulation data, verifying the violation of performance in the $n \mapsto V$ channel by controller $K_2$. Larger amplitudes than $0.2$ are not considered, as they result in a controller output that exceeds the $\pm 10 \mathrm{V}$ range of the disk's actuator.
 
Third, the results of the experiments with trapezoidal reference $r$ and noise $n(t) = 0.2 \sin(100t)$ are plotted together with the simulation data in Fig.~\ref{fig:time_sim_exp}, where the third row shows the controller output in simulation, and the bottom row the controller output in experiment. For the controller $K_1$ in Fig.~\ref{fig:time_sim_exp_K1}, all data from the experiment matches the simulation data closely. This is expected, since the voltages remain within the saturation bounds of the disk's actuator. Also, the agreement between simulation and experiment indicates that static friction, which is not modeled in~\eqref{eq:ode-disk}, does not seem to play a significant role in this situation. The latter is consistent with the hypothesis that static friction is responsible for the deviation at $a=0.1$ in Fig.~\ref{fig:sins_compare_exp}, and is not significant at larger amplitudes. For the controller $K_2$, the voltages produced by the controller exceed the $\pm 10\mathrm{V}$ range, and the effect of saturating the controller output results in a large deviation of the experimental data from the simulation data in Fig.~\ref{fig:time_sim_exp_K2}. The latter demonstrates the practical implication of the filter violation in the bottom right plot of Figs.~\ref{fig:bode_K2_static} and~\ref{fig:bode_K2_zf}; a performance violation in the $V$ channel can cause bad performance due to the actuator reaching its saturation bounds, which can be prevented if the controller satisfies the performance requirements (as $K_1$ does).

To conclude, as we have shown in this section, the differences between simulation and experiment are either very small, or they can be reasonably explained by unmodeled dynamics (e.g., static friction). Since the experimental results largely agree with the simulation results, and therefore also with the Bode diagram analysis, they validate our method in the sense that our predictions are not critically sensitive to small modelling errors. 

In~\eqref{eq:ode-disk}, all friction effects are identified via a viscous friction term~\cite{koelewijnPhysicalParameterEstimation2019}, and the voltage saturation is not modeled. In future work, we aim to refine our nonlinear Bode diagram analysis of the unbalanced disk setup by including the voltage saturation and static friction in the model~\eqref{eq:ode-disk}. Via multipliers that describe static friction as in~\cite{rantzerFrictionAnalysisBased2001}, we could verify that the deviation between the experimental RMS gain and the sinusoidal gain bound in Fig.~\ref{fig:sins_compare_exp} is indeed caused by static friction. The latter requires identifying a model for the unbalanced disk as in~\cite{koelewijnPhysicalParameterEstimation2019}, but now with an additional static friction term.

\section{Conclusion}\label{sec:conclusion}

This paper develops a Bode diagram for nonlinear and/or uncertain MIMO systems, which is efficiently computed via IQCs and the generalized KYP lemma. We define frequency-dependent gains for systems that preserve the period of the input: the harmonic, sinusoidal, and subharmonic gain. The harmonic gain is computed using inputs with a given period, the sinusoidal gain uses only sinusoidal inputs, and the subharmonic gain uses inputs with frequency content up to a given frequency. The harmonic and sinusoidal gains can be compared with describing function-based methods, and generalizations thereof, while the subharmonic gain provides a novel metric for the low-frequency performance of a nonlinear system. Plotting these gains as function of frequency for every input/output channel yields a Bode diagram for nonlinear systems. For multivariable systems, an extra column is added to the Bode diagram to capture the nonlinear effect of multiple nonzero input channels on one output channel. Moreover, input and output weighting filters from LTI mixed-sensitivity shaping can be used to specify performance. For systems that preserve the period of the input, we prove that the $L_2$-gain in the time-domain, which includes transients, is asymptotically equal to the RMS gain from periodic input to steady-state output. Then, we develop IQC-based tools to compute guaranteed upper bounds for these frequency-dependent gains. Here, the generalized KYP lemma is used to make these tools computationally tractable via semidefinite programming. 

We apply our proposed methods on the performance analysis of a controlled unbalanced disk. We show how the nonlinear Bode diagrams are computed in practice, and how they can be used to quantify the performance of a nonlinear control system. When a static nonlinearity is involved, we demonstrate that the integrator behavior of a controller in closed loop can be correctly quantified if Zames-Falb multipliers are used. We validate the performance description of our nonlinear Bode diagrams via simulations, and verify if the specifications from the weighting filters are met. All simulations are repeated as experiments on the setup, which shows that our method is not critically sensitive to model mismatches.

\section*{Declaration of AI Use}

Generative AI tools (Claude Opus 5.5) were used for the following parts of the work presented in this paper: proofreading of the manuscript for correctness and typos, software testing for the simulations and nonlinear Bode plot computations, and improving the scripts for plotting the Bode diagrams, simulation data and experimental data. All AI-generated content used has been reviewed by the authors, and the authors take full responsibility for the content of this paper.

\footnotesize
\bibliographystyle{IEEEtran} 
\bibliography{bibliography}

\color{black}

\end{document}